\documentclass[openany,12pt,a4paper,oneside,english,spanish,brazil]{abntex2}

\usepackage{amsmath}
\usepackage{amssymb}
\usepackage{amsthm}
\usepackage{bbm}
\usepackage{color}
\usepackage{float}
\usepackage[T1]{fontenc}
\usepackage{graphicx}
\usepackage{hyperref}
\usepackage[utf8]{inputenc}
\usepackage{orcidlink}
\usepackage{siunitx}
\usepackage{physics}
\usepackage{times}
\usepackage{xcolor}
\usepackage{comment}
\usepackage{siunitx}
\usepackage{titlesec}
\usepackage{setspace}
\usepackage{booktabs} 
\usepackage{siunitx}  
\usepackage{float}
\RenewCommandCopy\pqty\qty   
\AtBeginDocument{\RenewCommandCopy\qty\SI} 
\DeclareUnicodeCharacter{202F}{\,}

\usepackage{animate} 
\usepackage{times}	 
\usepackage[T1]{fontenc}		
\usepackage[utf8]{inputenc}		
\usepackage{lastpage}			
\usepackage{indentfirst}		
\usepackage[normalem]{ulem}     
\usepackage{color}				
\usepackage{graphicx}			
\usepackage{microtype} 			
\usepackage{leading}            
\usepackage{amsmath}
\usepackage{bbold}
\usepackage{amsfonts}
\usepackage{amssymb}
\usepackage{amsthm}
\usepackage{pdfpages}
\usepackage{psfrag}
\usepackage{float} 
\usepackage{caption}
\usepackage{subcaption} 
\makeatletter
\@cons\caption@subtypelist{{figure}}
\makeatother
\usepackage{adjustbox} 
\usepackage{diagbox} 
\usepackage{fancyhdr}
\usepackage{bookmark}
\usepackage{fr-longtable, ltcaption}
\usepackage{threeparttablex} 
\usepackage{hhline} 
\usepackage{booktabs}
\usepackage{physics}
\usepackage{hyperref}
\usepackage{cite} 
\usepackage[most]{tcolorbox} 
\usepackage{lmodern} 
\usepackage{xcolor}  
\usepackage[none]{hyphenat}

\usepackage{tocloft}
\usepackage{titletoc} 
\definecolor{mygrey}{rgb}{0.85, 0.85, 0.85}

\renewcommand{\sin}{\operatorname{sen}}

\usepackage{listings}
\definecolor{codegreen}{rgb}{0,0.6,0}
\definecolor{codegray}{rgb}{0.5,0.5,0.5}
\definecolor{codepurple}{rgb}{0.58,0,0.82}
\definecolor{backcolour}{rgb}{0.95,0.95,0.92}

\lstdefinestyle{mystyle}{
	backgroundcolor=\color{backcolour},   
	commentstyle=\color{codegreen},
	keywordstyle=\color{magenta},
	numberstyle=\tiny\color{codegray},
	stringstyle=\color{codepurple},
	basicstyle=\footnotesize,
	breakatwhitespace=false,         
	breaklines=true,                 
	captionpos=b,                    
	keepspaces=false,                 
	numbers=left,                    
	numbersep=5pt,                  
	showspaces=false,                
	showstringspaces=false,
	showtabs=false,                  
	tabsize=4
}

\usepackage{epstopdf}
\usepackage{lipsum}				

\usepackage[english,hyperpageref]{backref}	 

\renewcommand*{\backrefalt}[4]{
	\ifcase #1 %
		No citations in the text.%
	\or
		Cited on page #2.%
	\else
		Cited #1 times on pages #2.%
	\fi}%
\usepackage{pgf,tikz}
\usepackage{mathrsfs}
\usetikzlibrary{arrows}

\usepackage[num,abnt-doi=link]{abntex2cite}

\usepackage{comment}
\newcommand{\refanexo}[1]{\hyperref[#1]{Annex~\ref{#1}}} 

\setlrmarginsandblock{3cm}{2cm}{*}
\setulmarginsandblock{3cm}{2cm}{*}
\checkandfixthelayout

\definecolor{blue}{RGB}{41,5,195}

\makeatletter
\hypersetup{
   	pdfsubject={\imprimirpreambulo},
    pdfcreator={LaTeX with abnTeX2},
    pdfkeywords={abnt}{latex}{abntex}{abntex2}{academic work},
	colorlinks=true,       		
	linkcolor=black,          	
	citecolor=black,        		
	filecolor=black,      		
	urlcolor=black,	
	bookmarksdepth=4
}
\makeatother

\makeindex
\begin{document}
\selectlanguage{english}

\frenchspacing 
	\thispagestyle{empty}
	\noindent
	
	\begin{center}
	\textbf{UNIVERSIDADE ESTADUAL DE PONTA GROSSA}\\
	\textbf{SETOR DE CIÊNCIAS EXATAS E NATURAIS}\\
	\textbf{PROGRAMA DE PÓS-GRADUAÇÃO EM CIÊNCIAS}

    \vspace*{\fill}
	\textbf{EDUARDO KRONBAUER SOARES}\\  	
  	
  	\vspace*{3.5cm}
  	
	\textbf{SPIN CHAINS FOR QUANTUM INFORMATION PROCESSING}
	    
    	\end{center}
    	
		\vspace*{\fill}
\begin{center}
		\vfill 
	\textbf{\normalsize PONTA GROSSA}\\                               \par\vspace{4pt}
	\textbf{2026}
\end{center}
\newpage
\newpage
\begin{center}
{\textbf{EDUARDO KRONBAUER SOARES} }
\end{center}

\vfill
\vfill

\begin{center}  {\textbf{SPIN CHAINS FOR QUANTUM INFORMATION PROCESSING}} 
\end{center}

\vspace{0.5cm}

\hfill\begin{minipage}[H]{0.55\hsize}
					
                      Dissertation presented to the Graduate Program in Sciences,
    concentration area Physics, of the Universidade Estadual
de Ponta Grossa, in partial fulfillment of the requirements
for the degree of Master in Physics.\\\\
Advisor: Prof. Dr. Fabiano Manoel de Andrade.

					\end{minipage}
\vfill
\begin{center}
		\vfill 
	\textbf{\normalsize PONTA GROSSA}\\                               \par\vspace{4pt}
	\textbf{2026}
\end{center}


	\begin{agradecimentos}[ACKNOWLEDGMENTS]
\vspace{0.8cm}
I would like to express my gratitude to my family and friends for their support over the years. I am profoundly grateful to my advisor, Prof. Dr. Fabiano Manoel de Andrade, who has guided me since my first scientific project and helped me to not only understand Physics, but also to be sure of my love for it. I thank UEPG'S Quantum Physics and Quantum Information group (QPQI) for the productive discussions. I also wish to thank Prof. Dr. Gentil D. de Moraes Neto for his valuable help, particularly in understanding the formalism of open systems in the context of spin chains. I thank CAPES for the financial support. Finally, I acknowledge my perseverance through moments of difficulty.
\end{agradecimentos}

\begin{epigrafe}
\vspace*{\fill}
\begin{flushright}
\textit{\textquotedblleft
The understanding can intuit nothing, the senses can think nothing. \\ Only through their union can knowledge arise.\textquotedblright
\\(Immanuel Kant; Critique of Pure Reason, 1781)
}
\end{flushright}
\end{epigrafe}

\setlength{\absparsep}{18pt} 
\begin{resumo}[\textbf{ABSTRACT}]
Classical computation relies heavily on information manipulation. Each component of a hardware needs to communicate with others, and this is done by encoding information into strings of bits and application of logical operations. When dealing with quantum technologies, there arises a  new set of paradigms and devices, based on manipulations of qubits, the quantum analogues of conventional bits. This work investigates the generation and distribution of quantum entanglement, a uniquely non-classical correlation, across spin chains, which serve as promising platforms for quantum information processing. We systematically compare two distinct entanglement generation protocols: Protocol 1 (P1), based on alternating weak and strong couplings that create a band structure enabling an effective trimer-model approximation, and Protocol 2 (P2), which employs symmetric boundary couplings and virtual excitations to establish a direct effective interaction between the chain ends. Our results demonstrate that a protocol based on virtual excitations and optimized boundary couplings consistently outperforms its counterpart in speed, achieved entanglement, and robustness against fabrication imperfections and noise. Furthermore, by employing effective model reductions and open quantum systems techniques we provide a comprehensive framework for understanding the resilience of distributed entanglement in solid-state quantum devices. The characteristics of the virtual-coupling protocol highlight its potential for experimental implementation in scalable quantum technologies.

\textbf{Keywords:} Quantum Spin Chains; Entanglement Generation; XX Model; Quantum
Information Processing.
\end{resumo}
\setlength{\absparsep}{18pt} 
\begin{resumo}[\textbf{RESUMO}]
A computação clássica depende fortemente da manipulação de informação. Cada componente de um hardware precisa se comunicar com os outros, e isso é feito codificando informação em sequências de bits e aplicando operações lógicas. Ao lidar com tecnologias quânticas, surge um novo conjunto de paradigmas e dispositivos, baseados na manipulação de qubits, os análogos quânticos dos bits convencionais. Este trabalho investiga a geração e distribuição de emaranhamento quântico, uma correlação exclusivamente não clássica, em cadeias de spins, que servem como plataformas promissoras para o processamento de informação quântica.  Comparamos sistematicamente dois protocolos distintos de geração de emaranhamento: Protocolo 1 (P1), baseado em acoplamentos alternados fracos e fortes que criam uma estrutura de bandas permitindo uma aproximação efetiva pelo modelo de trios (trimer), e Protocolo 2 (P2), que emprega acoplamentos simétricos nas bordas e excitações virtuais para estabelecer uma interação efetiva direta entre as extremidades da cadeia.  Nossos resultados demonstram que o protocolo baseado em excitações virtuais e acoplamentos otimizados nas bordas supera consistentemente seu contraparte em termos de velocidade, emaranhamento alcançado e robustez contra imperfeições de fabricação e ruído. Além disso, ao empregar reduções de modelos efetivos e técnicas de sistemas quânticos abertos, fornecemos uma estrutura abrangente para compreender a resiliência do emaranhamento distribuído em dispositivos quânticos de estado sólido. As características do protocolo de acoplamento virtual destacam seu potencial para implementação experimental em tecnologias quânticas escaláveis.

\textbf{Palavras-chave:} Cadeias de Spins Quânticos; Geração de Emaranhamento; Modelo XX; Processamento de Informação Quântica.

\end{resumo}

\renewcommand\listfigurename{\normalsize \textbf{LIST OF FIGURES} \setlength{\afterchapskip}{\baselineskip}}
{
    \let\oldnumberline\numberline%
    \renewcommand{\numberline}{\figurename~\oldnumberline}
    \cftsetindents{figure}{0em}{2em}%
    \listoffigures*
}
\newpage
\renewcommand\listtablename{\normalsize \textbf{LIST OF TABLES}\setlength{\afterchapskip}{\baselineskip}}
{
    \let\oldnumberline\numberline%
    \renewcommand{\numberline}{\hspace{0cm} \tablename~\oldnumberline}%
    \cftsetindents{table}{0em}{2em}%
    \listoftables*%
    \cleardoublepage
}
\newpage

\renewcommand\listadesiglasname{\normalsize \textbf{LIST OF ABBREVIATIONS AND ACRONYMS}}
\begin{siglas}
\vspace{0.7 cm}
\item [P1] Protocol 1
\item[P2] Protocol 2 
\item[QM]  Quantum Mechanics
\item[QIT] Quantum Information Theory
\item[SC(s)] Spin Chain(s)
\item[VNE] Von Neumann Equation
\item[TCG] Time Coarse Graining
\item[LU] Local Unitary
\item[LME] Lindblad Master Equation 
  

\end{siglas}

\setlength{\cftchapternumwidth}{2em}
\renewcommand\contentsname{\normalsize \textbf{TABLE OF CONTENTS} \setlength{\afterchapskip}{\baselineskip}} %

\pdfbookmark[0]{TABLE OF CONTENTS}{toc}
\renewcommand*{\cftchapterafterpnum}{\vskip-\baselineskip\hspace*{2em}}
\cftsetindents{chapter}{0em}{2em}
\tableofcontents*
\cleardoublepage

\textual
\pagestyle{simple}

\chapter{INTRODUCTION}
\vspace{0.8cm} 
It was during the 17th century that Isaac Newton established what we now know as Physics \cite{newton1687principia}, referred to then as Natural Philosophy, providing a mathematical description of motion and gravitation. Newton's combination of math and observation encouraged the next generations of scientists and led to the systematic exploration of heat, work, temperature, and energy, the area we understand as Thermodynamics today. The new discipline was motivated in large part by the need to improve steam engines, which, unlike the traditional water-wheels, did not need rivers and geography to drive them and could be the real movers of industry. In this context, Sadi Carnot and Rudolf Clausius, utilizing physical intuition and philosophical investigation into the nature of energy, developed the theory and helped increase the efficiency and productivity of these machines.

Steam power was unparalleled in efficiency and flexibility, leading to the First Industrial Revolution and dramatically changing the course of human history. This period, spanning the late eighteenth and early nineteenth centuries, marked the transition from rural communities to industrial economies through mechanization, factories, and new types of work \cite{hanlon2020block}. 

The Second Industrial Revolution of the late nineteenth and early twentieth centuries was powered by electricity and electromagnetism. Maxwell's equations provided the theory for this revolution, enabling electric lighting and telecommunication but also laying the foundations of Modern Physics \cite{bowler2010making}. 

The Third Industrial Revolution (also called \textit{the Information Age})\cite{castells1996information}, which occurred during the mid-twentieth century, was shaped by the advent of digital technology. The development of the transistor, whose origin can be traced to Quantum Mechanics (QM), and the evolution of Information Theory through the works of Claude Shannon and John von Neumann revolutionized computation and communication. Computers, satellites, and the internet revolutionized the frontiers of knowledge and connectivity.

Today, humanity is in the middle of the Fourth Industrial Revolution, characterized by the confluence of Quantum Information, Artificial Intelligence, Nanotechnology, and Biotechnology \cite{schwab2024fourth}. In this context, Quantum Information research has emerged to take on crucial importance. Algorithms such as Shor's algorithm are expected to render classical cryptography protocols obsolete upon being realized on sufficiently powerful quantum computers \cite{Bagirovs_2024}. At the same time, experimental triumphs such as quantum teleportation, already achieved over kilometer-scale distances \cite{PRXQuantum.1.020317}, signal the accelerating pace of advancement in the field. At the center of all such ideas are Quantum Information Theory (QIT) and, in particular, entanglement as a resource. 

And in these days, just as Clausius, Carnot, Maxwell, Shannon, and others developed theoretical frameworks that pushed the technological frontiers of their time, scientists now find themselves doing the same through QIT. A deeper understanding of quantum information not only enriches our understanding of physics itself but also drives technological innovation. And these innovations have already begun to transform—and will continue to transform—the world and the way in which we live.

It is in this context that spin chains (SCs) emerge as candidates for quantum technological applications. They are natural and flexible systems not only for generating but also for distributing and transmitting quantum entanglement across sites within a solid-state platform \cite{Bose2003, Christandl2004, Yao2011, Subrahmanyam2004,Kay2010}. Their natural compatibility with direct integration into solid-state implementations makes them promising candidates for chip-based architectures, including those envisioned for large-scale quantum computers. 

In this work, the ability to rapidly generate entangled states while maintaining robustness against fabrication imperfections and environmental noise is of foremost importance. Establishing protocols capable of delivering high-fidelity entanglement distribution under realistic, noisy conditions is therefore as much a pragmatic step toward the advancement of contemporary technologies as it is a theoretical challenge.

We compare two entanglement generation protocols based on XX-type SCs: Protocol 1 (P1), where alternating weak and strong couplings generate quantum correlations at the chain edges, and Protocol 2 (P2), which employs symmetric couplings at both ends to enhance the transport speed and facilitate the buildup of quantum entanglement. We systematically investigate the entanglement dynamics for spins $s = \frac{1}{2}, \text{ }1, \text{ and } \frac{3}{2}$ under ideal conditions. For the $s=1/2$ case, we also examine non-ideal cases. Our analysis is primarily based on extensive numerical simulations. 

This dissertation is organized as follows. Chapter 2, presents the theoretical foundations of Quantum Information, including the mathematical formalism of Hilbert spaces, density matrices, and others. Chapter 3 introduces the spin chain models used in this work, describes in detail the entanglement generation protocols P1 and P2, and discusses why these protocols are viable candidates for efficient entanglement generation. Chapter 4 investigates the effects of static disorder, both diagonal and off-diagonal, on the entanglement dynamics. Chapter 5 extends the analysis to open quantum systems, incorporating environmental decoherence via Lindblad master equation (LME). Finally, Chapter 6 summarizes the main results and discusses the broader implications of our findings for the development of robust and scalable quantum technologies.


\chapter{QUANTUM INFORMATION THEORY} 
\label{chap:info_thoery}

\vspace{0.8cm}
The origins of QM date back to the early 20th century, when classical theories failed to explain several physical phenomena. Black-body radiation \cite{planck1900theory}, the photoelectric effect \cite{einstein1905heuristic}, and atomic spectra \cite{bohr2025constitution} are examples that defied classical explanations. Max Planck, who introduced the quantization of energy in 1900, Albert Einstein, who applied this concept to explain the photoelectric effect in 1905 and Louis de Broglie, who proposed wave–particle duality, laid the foundations of quantum theory. Afterwards, QM advanced further through the contributions of Niels Bohr, Werner Heisenberg, Erwin Schrödinger, Paul Dirac and others.

By the mid-20th century, QM had become an established branch of physics, transforming our understanding of nature at the microscopic level. This period is known as the first quantum revolution. Today, we are witnessing the second quantum revolution, in which the principles developed during the first revolution are being used to design new technologies based on QM. In his seminal work on the mathematical foundations of quantum mechanics \cite{von1947fondements}, von Neumann introduced the concept of quantum entropy, now known as the von Neumann entropy, as a natural extension of the classical entropy to quantum statistical ensembles. This quantity provides a fundamental measure of mixedness and information content in quantum states, and it later became one of the central concepts of quantum information theory.

During the second half of the 20th century, the field underwent remarkable development, yielding a series of breakthroughs that established a solid basis for emerging areas such as quantum computing and quantum technologies. Notable examples of quantum algorithms rooted in QIT include Shor’s algorithm for efficient integer factorization \cite{Shor1997}, the Deutsch–Jozsa algorithm for solving black-box query problems \cite{Deutsch1992}, and Grover’s search algorithm, which provides a quadratic speedup for unstructured search tasks \cite{Grover1997}. On the technological side, protocols such as quantum teleportation can be experimentally realized today only because of the robust theoretical foundations provided by QIT \cite{Bennett1993}.

In the following sections, we outline the fundamental principles of QM, followed by the essential definitions and concepts in QIT that underpin our exploration.

\section{QUANTUM MECHANICS}
Quantum mechanics is a theory formulated in terms of a set of postulates. 
These postulates are empirically motivated assumptions that, together with the 
underlying mathematical framework, allow one to derive the predictions and theorems 
of the theory. The postulates of quantum mechanics may be stated as follows:

\begin{itemize}
    \item \textbf{State postulate:}  
    The state of a quantum system is completely specified by a complex wave function 
    $\psi(\mathbf{r},t)$, or equivalently by a state vector $|\psi(t)\rangle$ in a complex 
    Hilbert space. All measurable information about the system is contained in this state.

    \item \textbf{Observables postulate:}  
    Every physical observable $A$ is represented by a linear Hermitian operator 
    $\hat{A}$ acting on the state space of the system.

    \item \textbf{Measurement postulate:}  
    The only possible outcomes of a measurement of an observable $A$ are the eigenvalues 
    $a_n$ of the corresponding operator $\hat{A}$. The probability of obtaining the 
    outcome $a_n$ when the system is in the state $|\psi\rangle$ is given by
    \[
        P(a_n) = |\langle a_n | \psi \rangle|^2 ,
    \]
    where $|a_n\rangle$ is the eigenstate associated with $a_n$.

    \item \textbf{State reduction postulate:}  
    Immediately after a measurement yielding the value $a_n$, the state of the system 
    collapses to the corresponding eigenstate $|a_n\rangle$.

    \item \textbf{Time evolution postulate:}  
    The time evolution of the state vector $|\psi(t)\rangle$ is governed by the 
    Schrödinger equation
    \[
        i\hbar \frac{\partial}{\partial t} |\psi(t)\rangle = \hat{H} |\psi(t)\rangle ,
    \]
    where $\hat{H}$ is the Hamiltonian operator of the system.

    \item \textbf{Expectation value postulate:}  
    The expectation value of an observable $A$ in the state $|\psi\rangle$ is given by
    \[
        \langle A \rangle = \langle \psi | \hat{A} | \psi \rangle .
    \]
\end{itemize}

Now, we define the properties of the vector space used in QM, known as a Hilbert space \cite{halmos_1951,Juliane2012}. A vector space is a collection of elements, called vectors, that is closed under addition and multiplication by scalars. A Hilbert space \( \mathcal{H} \) is a vector space equipped with an inner product \( \langle \psi | \phi \rangle \) for any \( \{ \ket{ \psi }, \ket\phi \} \in \mathcal{H} \), such that the norm defined by
\begin{align}
\|\phi\| = \sqrt{\langle \phi | \phi \rangle},
\end{align}
makes \( \mathcal{H} \) a complete metric space (see Appendix \ref{apendB}).

The dimensions of the complete Hilbert spaces in this dissertation are always \( 2^N \) because we are dealing with systems containing a finite number \( N \) of distinguishable spin-half particles in a chain. 
This allows us to define \( \ket{\phi}, \ket{\psi} \in \mathcal{H} \) as
\begin{align}
\ket{\phi} &= \sum_{i=1}^{2^N} c_i \ket{\phi_i}, \\
\ket{\psi} &= \sum_{i=1}^{2^N} d_i \ket{\psi_i},
\end{align}
where \( \{\ket{\phi_i}\} \) and \( \{\ket{\psi_i}\} \) are orthonormal bases of \( \mathcal{H} \) and the inner product of \( \ket{\phi} \) and \( \ket{\psi} \) is given by
\begin{align}
\bra{\phi}\ket{\psi} = \sum_{i=1}^{2^N} c_i^* d_i.
\end{align} 
A Hilbert space of dimension \(2^N\) is constructed as the tensor product of the individual Hilbert spaces,
\begin{align}
\mathcal{H} = \mathbb{C}^{2^N} = \bigotimes_{i=1}^N \mathbb{C}^2.
\end{align}
The Hilbert space of each spin-half has a basis known as the computational basis, denoted by \( \{\ket{0}, \ket{1}\} \), where
\begin{align}
\ket{0} = \begin{pmatrix} 1 \\ 0 \end{pmatrix}, \quad \ket{1} = \begin{pmatrix} 0 \\ 1 \end{pmatrix}.
\end{align}

According to the postulates of QM, the state of an isolated quantum system at a fixed instant of time \( t \) is described by a unit vector in the Hilbert space \(\mathcal{H} \). For systems that are not pure, we must introduce the density operator formalism \cite{nielsen_chuang_2010}. This formulation is mathematically equivalent to the state vector approach, but provides a more convenient framework for describing systems in which classical probabilities are introduced (e.g., due to decoherence).

To define the density operator, consider a quantum system that can be in one of a set of states \( \{\ket{\psi_i}\} \), each with a classical probability \( p_i \). The collection \( \{p_i, \ket{\psi_i}\} \) is called an ensemble of pure states. The density operator \( \rho \) is then defined as
\begin{align}
\rho \equiv \sum_i p_i \ketbra{\psi_i} {\psi_i}.
\end{align}
All the postulates of QM can be reformulated in terms of the density operator.  

The framework of density operators provides a natural way to compute expectation values. Since both full and reduced density matrices contain all accessible information about a quantum system or subsystem, expectation values of observables follow directly from them. For an observable \(O\) and a quantum state described by a density matrix \(\rho\), the expectation value is obtained through the trace operation \cite{cohen1991quantum},
\begin{align}
\langle O \rangle = \mathrm{Tr}(\rho O).
\end{align}
The trace of an operator acting on a Hilbert space $\mathcal{H}$ is defined as the sum of its diagonal matrix elements in any orthonormal basis $\{|i\rangle\}$. In the case of composite systems, if the observable acts only on subsystem \(A\), this expression reduces to
\begin{align}
\langle A \rangle = \mathrm{Tr}_A(\rho_A O),
\end{align}
where \(\rho_A\) is the reduced density matrix obtained via the partial trace. The partial trace of an operator over subsystem $A$ is obtained by summing over the matrix elements corresponding to an orthonormal basis of $\mathcal{H}_A$. Thus, expectation values naturally fit within the density operator formalism, reinforcing its central role in describing both global and local properties of quantum systems.

Since accessing information about a quantum system ultimately relies on quantities such as these expectation values, it becomes crucial to understand how the density matrix evolves in time. The Schrödinger equation cannot be applied directly, as it governs only the dynamics of pure state vectors. Instead, the coherent evolution of a density matrix is described by the von Neumann equation (VNE). For a closed quantum system with Hamiltonian $H$, the VNE takes the form,
\begin{align}
\frac{d\rho(t)}{dt} = -\frac{i}{\hbar} [H, \rho(t)],
\end{align}
where $\rho(t)$ is the density operator. This equation preserves the fundamental properties of the density matrix (trace, positivity, and hermiticity), ensuring consistency with the probabilistic interpretation of QM. Unlike the Schrödinger equation, which applies only to pure states, the VNE remains valid for any quantum state, whether pure or mixed.

The VNE is particularly important in QIT, where decoherence, entanglement, and measurement processes often require the density matrix formalism. In later sections, we will see how the VNE serves as a cornerstone for analyzing SCs and quantum information processing, especially in scenarios where pure-state descriptions are insufficient.

Another important aspect of the formalism is the way operators act on composite quantum systems, which follows directly from the tensor-product structure of the underlying Hilbert spaces. Consider two distinct quantum systems \(A\) and \(B\) with associated Hilbert spaces \(\mathcal{H}_A\) and \(\mathcal{H}_B\). When describing the combined system on \(\mathcal{H}_{AB} = \mathcal{H}_A \otimes \mathcal{H}_B\), operators acting on individual subsystems must be embedded into the composite space in a manner that preserves both the algebraic structure and the physical interpretation of their action.

A more general type of transformation that naturally extends the operator formalism in QM is the \textit{superoperator}. While ordinary operators act on state vectors within a Hilbert space, superoperators act on operators themselves, mapping one operator to another. Formally, a superoperator $\tilde{O}$ acts as
\begin{align}
    \tilde{O}(A) = B,
\end{align}
and may admit eigenoperators satisfying
\begin{align}
    \tilde{O}(A) = \lambda A.
\end{align}
Superoperators provide a compact way to describe processes such as quantum channels, dissipative dynamics, and general state transformations, playing a central role in the theory of open quantum systems.
\subsection{Quantum States}

Quantum states are fundamental to the study of quantum systems. They contain all the accessible information about a system's state. Thus, it is necessary to highlight some special cases of quantum states that will appear throughout this dissertation:

\begin{itemize}
    \item \textbf{Pure States}: A state is pure if and only if \(\text{Tr}(\rho^2) = 1\). A pure state represents a situation where we have complete knowledge of the quantum state, i.e., there is no classical uncertainty associated with the system's state.

    \item \textbf{Mixed States}: If a state is not pure, it is mixed. This means that \(\text{Tr}(\rho^2) < 1\). Mixed states arise when there is classical uncertainty about the system's state. Some states can be "more mixed" than others, indicating a higher degree of classical uncertainty. 

    \item \textbf{Product States}: Consider two Hilbert spaces \(\mathcal{H}_A\) and \(\mathcal{H}_B\). If \(\ket{\psi_A} \in \mathcal{H}_A\) and \(\ket{\psi_B} \in \mathcal{H}_B\), we can construct the product state of these states as
    \[
    \ket{\psi} = \ket{\psi_A} \otimes \ket{\psi_B}.
    \]
    Product states describe quantum systems composed of independent subsystems. They are particularly useful for describing systems such as SCs, which require multiple interacting Hilbert spaces for their proper representation.

    \item \textbf{Entangled States}: If a state describing systems \(A\) and \(B\), associated with Hilbert spaces \(\mathcal{H}_A\) and \(\mathcal{H}_B\) respectively, cannot be written as a product state, it is said to be entangled. Entangled states exhibit correlations between subsystems that cannot be described classically. Entangled states are extremely common in real life. The two electrons of a helium
atom in the ground state, for example, have their spins entangled, and indeed any
two quantum particles that are interacting with each other will most likely be entangled \cite{wootters1998quantum}. 
\end{itemize}

Entangled states are of great importance in QIT, as they represent a crucial resource for many quantum computation protocols \cite{horodecki2009quantum}. For example, Shor's and Grover's algorithms, mentioned in the introduction of this chapter, rely on entanglement at some stage of their operation. These algorithms demonstrate how entanglement can be harnessed to achieve computational advantages over classical methods. 
\section{FROM BITS TO QUBITS}

We now provide a concise introduction to the foundational concepts of classical Information Theory and QIT. We begin by discussing Shannon entropy, a fundamental quantity in classical information theory, and its role in defining the classical bit. We then extend these ideas to QIT, establishing a bridge between classical and quantum information.

\subsection{Shannon Entropy}

In 1948, while working at Bell Telephone Laboratories, Claude Shannon sought to quantify the loss of information in phone-line signals. This led to the development of Shannon entropy, a measure that became fundamental to Information Theory. Interestingly, Shannon wasn't initially aware that his equation resembled Boltzmann's entropy. It wasn’t until a conversation with  John von Neumann that Shannon realized the similarity, and von Neumann suggested the name entropy for Shannon's discovery \cite{CoverThomas2006}.

To understand the Shannon entropy, we first need to define a \textit{bit}, the most basic unit of classical information. A bit can take one of two values, \(0\) or \(1\). All forms of information can be encoded as strings of bits. Before we dive into Shannon entropy, let's first explore the concept of \textit{surprisal}, or the surprise associated with an outcome. Let $X$ denote a random variable, the surprisal associated with one specific outcome \(x \) of $X$ is given by,
\begin{align}
i(x) = -\log_2 \left[p(x)\right].
\end{align}
Here, \(p(x)\) is the probability of outcome \(x\). The surprisal measures how much information we gain from an outcome: outcomes with low probability yield high surprisal, meaning that the information gain is high when measuring $X$ and obtaining that specific outcome. This captures the idea that the more unexpected an outcome is, the more information we obtain by measuring it.

Now, consider that the random variable \(X\) can take on several possible values. One may ask: what is the average amount of information we gain from measuring \(X\)? This leads to the Shannon entropy,
\begin{align}
H(X) = - \sum_{i=1}^n p_i \log_2 p_i
\end{align}
where \(p_i\) is the probability of outcome \(i\), and the entropy is measured in bits, since the logarithm is base 2. Shannon entropy quantifies the average surprisal of the outcomes of a random variable.
\begin{figure}[h]
    \centering
    \caption{
    The biased coin has a 99\% chance of landing yellow and a 1\% chance of landing red. As a result, when an observer measures it on step 3, it provides minimal information, since the outcome is nearly certain.}
    \includegraphics[width=0.7\linewidth]{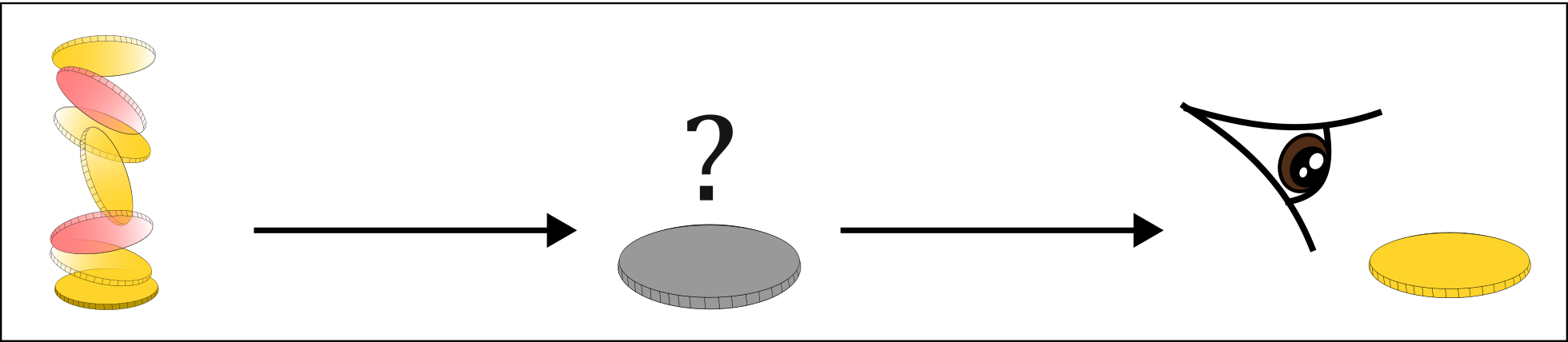}
    \caption*{{\small Source: The author.} }
    \label{fig:eyecoin}
\end{figure}
To better understand this concept, let's take a look at an example. Suppose we have a biased coin with two sides, one red and one yellow (see Figure \ref{fig:eyecoin}). Suppose that \(P(\text{yellow})=0.99\) and \(P(\text{red})=0.01\), the Shannon entropy of the coin is
\[
H(X) = - \left( 0.99 \log_2(0.99) + 0.01 \log_2(0.01) \right) \approx 0.0807 \text{ bits.}
\]
This low entropy indicates that, because one of the outcomes is almost certain, very little information is gained, on average, by measuring the coin. In contrast, consider a fair coin also with a red and a yellow side, but now the probabilities of yellow and red are both \(P(\text{yellow})=P(\text{red})=0.5\). The entropy of this fair coin is:
\[
H(X) = - \left( 0.5 \log_2(0.5) + 0.5 \log_2(0.5) \right) = 1 \text{ bit.}
\]
Here, because the outcomes are equally likely, the entropy is maximized, meaning the system is most uncertain, and, on average, we gain the most information from measuring it.

\begin{figure}[h]
    \centering
    \caption{Shannon entropy $H(X)$ of a binary random variable as a function of the probability $p(x)$.
The entropy is minimal when the distribution is fully biased ($p = 0$ or $p = 1$) and
reaches its maximum at the uniform distribution $p = 1/2$.
}
    \includegraphics[width=0.7\linewidth]{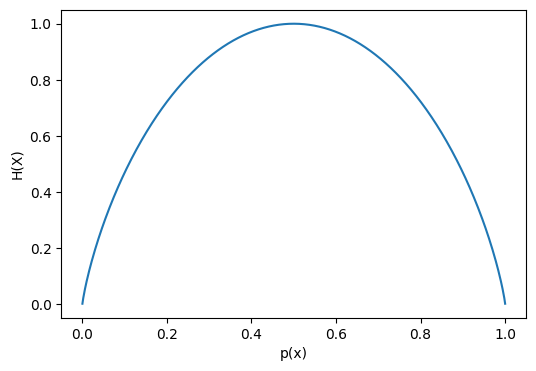}
    \caption*{{\small Source: The author.} }
    \label{fig:shannon}
\end{figure}

So, we can see that the Shannon entropy of a binary variable is lowest when the probability is concentrated on a single outcome, making the system more predictable and reducing the average informational surprise. Conversely, as the probability distribution becomes more balanced between the two possible outcomes, the entropy increases and reaches its maximum at $p = 1/2$ (see Fig. \ref{fig:shannon}). This behavior highlights that Shannon entropy quantifies the average surprise associated with the outcome of the system.

Some important properties of Shannon's entropy are:
\begin{itemize}
    \item \textbf{Maximized entropy}: When all outcomes are equally likely, the entropy reaches its maximum value. For a system with \(d\) possible outcomes, the maximum entropy is \( H(X) = \log_2 d \).
    \item \textbf{Minimized entropy}: If one outcome is certain (i.e., the probability of one outcome is 1), the entropy is 0, reflecting no uncertainty.
    \item \textbf{Additivity}: For independent random variables, the total entropy is the sum of the individual entropies.
\end{itemize}

In QIT, the analogous unit of the bit is the quantum bit or \textit{qubit}. However, qubits exhibit fundamentally different behavior due to the quantum mechanical principle of superposition.

A qubit represents a two-level quantum system with basis states $\ket{0}$ and $\ket{1}$. Unlike classical bits, qubits can exist in any coherent superposition of these basis states,
\[
\ket{\psi} = \alpha\ket{0} + \beta\ket{1},
\]
where $\alpha, \beta \in \mathbb{C}$ are probability amplitudes satisfying the normalization condition $|\alpha|^2 + |\beta|^2 = 1$. This superposition principle allows a qubit to occupy a continuous range of states between $\ket{0}$ and $\ket{1}$, in contrast to the strictly binary nature of classical bits.

The quantum state of a qubit can be geometrically represented on the Bloch sphere (see Appendix \ref{apendB}), where the south and north poles correspond to the basis states $\ket{0}$ and $\ket{1}$, and all other points on the sphere's surface represent valid superposition states. This visualization underscores the infinite continuum of possible qubit states, a stark departure from the two discrete states available to classical bits.
\begin{figure}[h]
    \centering
     \caption{Representation of the $\ket{\psi}=\frac{1}{\sqrt
    2}(\ket{0}+\ket{1})$ state in the Bloch sphere.}
    \includegraphics[width=0.5\linewidth]{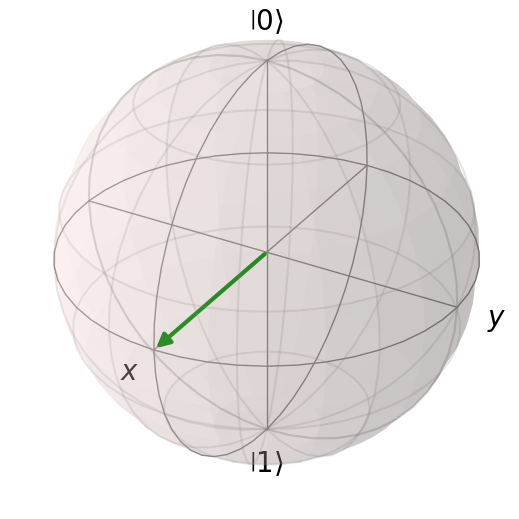}
    \caption*{{\small Source: The author.} }
    \label{fig:bloch}
\end{figure}

For a quantum state described by a density matrix $\rho$, the von Neumann entropy provides a measure of average surprisal, analogous to Shannon entropy in classical systems. It is defined as
\begin{align}
S(\rho) = -\text{Tr}(\rho \log_2 \rho).
\label{eq:von}
\end{align}
Von Neumann entropy quantifies the surprise associated with a quantum system. It increases as the system's classical probabilities become more spread out across possible basis states, indicating that it is harder to make a prediction about what outcome will result from a measurement, characterizing increase in the average surprise.

Some key properties of von Neumann entropy include:
\begin{itemize}[leftmargin=*]
    \item \textbf{Pure states}: If $\rho = \ketbra{\psi}$, then $S(\rho) = 0$ (no surprisal).
    \item \textbf{Maximally mixed states}: For $\rho = \mathbbm{I}/D$ (e.g., a mixed state with equal classical probabilities), where $D$ is the dimension of the Hilbert space, the von Neumann entropy is $S(\rho) = \log_2 D$ bits.
\end{itemize}

The von Neumann entropy plays a crucial role in quantifying surprisal in quantum systems, mirroring the role of Shannon entropy in classical Information Theory. Their foundational contributions were indispensable to the development of QIT and therefore are essential for a good understanding of our work and its significance. 

\section{MEASURES OF INFORMATION}

Having established the fundamental units and foundational quantities of information theory, we now focus on developing quantitative measures for characterizing quantum systems. This framework is essential for evaluating SCs as potential platforms for quantum information processing. Our primary objective in this work is to assess their capacity for generating entangled state pairs under various physical conditions. 

To achieve this, we require precise measures that can quantify several critical properties: the degree of entanglement between subsystems, the fidelity of state transfer through the chain, and the robustness of information preservation under environmental interactions. These quantitative tools will enable systematic evaluation of SC performance in quantum information protocols.

\subsection{Negativity}

Among the various entanglement quantifiers, negativity stands out as one of the most computationally tractable and algebraically straightforward measures \cite{Zyczkowski:1998yd}. Its operational significance becomes clear when examining its mathematical formulation,

\begin{equation}
\mathcal{N}(\rho_{AB}) = \frac{\|\rho_{AB}^{T_A}\|_1 - 1}{2}
\label{eq:neg}
\end{equation}
where $\rho_{AB}^{T_A}$ denotes the partial transpose of the bipartite density matrix $\rho_{AB}$ with respect to subsystem $A$, and $\|\cdot\|_1$ represents the trace norm operation.

 Negativity utilizes the distinctive behavior of separable versus entangled states under partial transposition. Consider the general density matrix expressed in the computational basis,
\begin{equation}
\rho_{AB} = \sum_{i,j;k,l} \rho_{i,j;k,l} \ketbra{i}{j} \otimes \ketbra{k}{l}.
\end{equation}
The partial transpose operation with respect to subsystem $A$ exchanges the bra and ket indices,
\begin{equation}
\rho^{T_A}_{AB} = \sum_{i,j;k,l} \rho_{i,j;k,l} \ketbra{j}{i} \otimes \ketbra{k}{l}.
\end{equation}
For separable states, which admit the decomposition,
\begin{equation}
\rho_{AB} = \sum_i p_i [\rho_A^i \otimes \rho_B^i],
\end{equation}
the partial transpose yields,
\begin{equation}
\rho^{T_A}_{AB} = \sum_i p_i [(\rho_A^i)^T \otimes \rho_B^i].
\end{equation}
Since the transpose preserves positivity for any valid density matrix $\rho_A$, the partial transpose of a separable state remains positive semidefinite. This property fails for entangled states, which cannot be expressed in a separable form.

The trace norm for a Hermitian operator reduces to:
\begin{equation}
\|\rho_{AB}^{T_A}\|_1 = \text{Tr}\sqrt{(\rho_{AB}^{T_A})^2} = \sum_i |\lambda_i|,
\end{equation}
where $\{\lambda_i\}$ are the eigenvalues of $\rho_{AB}^{T_A}$. For separable states, this norm equals unity due to trace preservation, yielding zero negativity. For entangled states, the deviation from zero quantifies the degree of entanglement, with larger deviations indicating a higher degree of entanglement in the system.

The negativity thus serves as a faithful entanglement witness, measuring the distance between the given state and the convex set of separable states. The maximum value of the negativity, $\mathcal{N}_{\text{max}}$, depends on the dimension of the quantum system. We therefore work with the normalized negativity, 
$$\frac{\mathcal{N}(\rho_{AB}(t))}{\mathcal{N}_{\text{max}}},$$ 
to establish a universal scale for comparing entanglement across different dimensions.

\subsection{Classical and Quantum Fidelity}

The concept of fidelity originates classically from the Bhattacharyya coefficient, which quantifies the similarity between two probability distributions \cite{nielsen_chuang_2010}. For discrete probability distributions $P = (p_1, \ldots, p_n)$ and $Q = (q_1, \ldots, q_n)$, it is defined as:
\begin{equation}
F_{\text{class}}(P,Q) = \sum_{i=1}^n \sqrt{p_i q_i}
\end{equation}
This measure exhibits several key properties:
\begin{itemize}
    \item \textbf{Identity}: $F_{\text{class}}(P,P) = 1$ when the distributions are identical.
    \item \textbf{Orthogonality}: $F_{\text{class}}(P,Q) = 0$ when the supports are disjoint ($p_i q_i = 0$ for all $i$).
    \item \textbf{Monotonicity}: The value increases as the distributions become more similar.
\end{itemize}

\begin{figure}[h]
    \centering
    \caption{
    An example of the calculation of the Bhattacharyya coefficient for two gaussian probability distributions.}
    \includegraphics[width=0.7\linewidth]{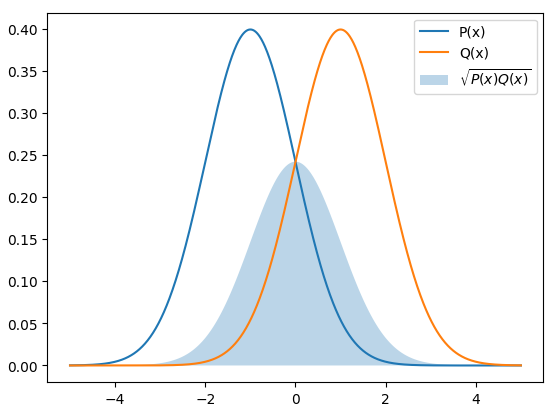}
    \caption*{{\small Source: The author.} }
    \label{fig:batta}
\end{figure}

The coefficient fundamentally relies on the calculation of $\sqrt{p_i q_i}$, which achieves its maximum when $p_i = q_i$ and decreases as the difference $|p_i - q_i|$ grows. This property makes it particularly sensitive to both the overlap and relative shapes of the distributions (for example, see Figure \ref{fig:batta}).

Extending this concept to quantum states requires careful consideration of non-commutativity and mixed states. The naive approach $\text{Tr}(\rho\sigma)$ is flawed; it is non-symmetric and does not equal 1 for identical mixed states, since $\text{Tr}(\rho^2) < 1$ for any mixed $\rho$.

The quantum fidelity between density matrices $\rho$ and $\sigma$ is properly defined as \cite{VidalWerner2002}:

\begin{equation}
F(\rho, \sigma) = \text{Tr}\left( \sqrt{\rho^{1/2} \sigma \rho^{1/2}} \right)
\end{equation}
This formulation preserves the essential characteristics of the classical fidelity while respecting quantum mechanical principles:

\begin{enumerate}
    \item \textbf{Normalization}: $F(\rho,\rho) = 1$.
    \item \textbf{Symmetry}: $F(\rho,\sigma) = F(\sigma,\rho)$.
    \item \textbf{Unitary invariance}: $F(U\rho U^\dagger, U\sigma U^\dagger) = F(\rho,\sigma)$.
    \item \textbf{Consistency}: For commuting $\rho$ and $\sigma$, reduces to classical fidelity.
\end{enumerate}

This measure proves particularly valuable in quantum information processing for evaluating state preparation accuracy, quantum channel performance, and entanglement verification, all crucial for assessing SC dynamics in our investigation \cite{Nielsen2010,Watrous2018,Horodecki2009}.

Having established the fundamental concepts of QIT, we can now focus on applying this framework to SCs. Thus in the next chapter, we introduce the physical model of SCs, show the relevant Hamiltonians, and discuss how these systems can be engineered to transfer, process, and generate quantum entanglement. The theoretical tools developed in the present chapter, especially reduced states, entanglement quantification, and dynamical evolution via the VNE, will serve as the basis for analyzing their performance throughout this dissertation.

\chapter{SPIN CHAINS}

\label{chap:spin_chains}
\vspace{0.8cm}

In early QM, spin was discovered to be an essential quantum number for describing particles. A landmark demonstration of this property came with the Stern-Gerlach experiment \cite{Gerlach1922}, where a beam of silver atoms passed through an inhomogeneous magnetic field and was deflected into a discrete spectrum, either up or down. This phenomenon could only be explained by introducing the concept of spin as an intrinsic property of electrons.

The theoretical foundation for spin was significantly advanced by two students advised by Paul Ehrenfest, George Uhlenbeck and Samuel Goudsmit, who hypothesized that electrons possess an intrinsic spin \cite{Uhlenbeck1925}. This groundbreaking idea provided a natural explanation for the observed splitting of atomic spectra and the behavior of electrons in magnetic fields.

Shortly after, in 1926, Paul Dirac and Enrico Fermi independently developed the full quantum statistics for electrons, now known as Fermi--Dirac statistics. These statistics describe particles with half-integer spin (fermions) and obey the Pauli exclusion principle, which states that no two fermions can occupy the same quantum state simultaneously. Fermi--Dirac statistics proved to be highly effective in describing a wide range of phenomena, including the collapse of stars into white dwarfs \cite{Fowler1926} and the behavior of electrons in metals \cite{Sommerfeld1927}.

In this context, Werner Heisenberg developed his theory of exchange interaction \cite{Heisenberg1928}, a quantum mechanical effect arising from the symmetry requirements of the wavefunction for identical particles. This interaction, fundamentally linked to the Pauli exclusion principle, describes how the spins of particles influence their mutual behavior. Heisenberg’s work laid the foundation for understanding magnetic interactions in solids and the alignment of spins in ferromagnetic materials.

\section{HAMILTONIAN}

An N-site distinguishable spin-1/2 chain with nearest-neighbor interactions and open boundary conditions (i.e. the $N$th spin doesn't couple with the first one) can be described by the following general Heisenberg Hamiltonian

\begin{align}
{H_{HM}} = \frac{1}{2} \sum_{i=1}^{N-1} J_{i,i+1} \left[ (1+\lambda)S_x^iS_x^{i+1} + (1-\lambda)S_y^iS_y^{i+1} + \Gamma S_z^iS_z^{i+1} \right] + \sum_{i=1}^N \epsilon_iS_z^i + \sum^N_{i=1}B_iS^i_z
\label{eq:hamiltonian}
\end{align}

The on-site energy, $\epsilon_i$, represents the energy of an individual spin and is considered homogeneous if it is identical across all sites. This parameter effectively shifts the energy gap between the two fundamental spin states. The coupling energy governs the strength of the interaction between neighboring sites, $J_{i,i+1}$. The symmetry, or anisotropy, of this spin-spin interaction defines several key models: the isotropic Heisenberg (XXX) model ($\lambda = 0, \Gamma = 1$); the XXZ model ($\lambda = 0, \Gamma \neq 1$), whose special case with $\Gamma = 0$ is known as the XX (or YY) model. For this chapter, we set $\epsilon_i = 0$ to simplify the system, noting that this parameter will later be used to introduce disorder. The final term represents a local magnetic field of strength $B_i$ acting at site $i$. 

The SCs configurations analyzed in this dissertation fall into two distinct
classes, referred to as P1 and P2. P1 is based
on a class of dimerized spin chains that have been previously studied in the
literature as platforms for entanglement
generation \cite{Estarellas2018}. Its properties and performance are therefore well understood under
idealized conditions.

By contrast, P2 is introduced in this work as an alternative architecture specifically designed to generate faster and highly entangled states, while also improving robustness against imperfections such as static disorder and environmental perturbations. While inspired by
existing state-transfer schemes, P2 exploits a different dynamical mechanism,
leading to faster entanglement generation and enhanced resilience, as will be
demonstrated throughout this dissertation.
\subsection{Pauli Exclusion Principle and the Heisenberg Model}

To elucidate the fundamental behavior of the Heisenberg Hamiltonian, also called the Heisenberg Model, we first analyze the case of two interacting spins. This approach provides a good intuition behind the model, and explains the general $N$-spin system with nearest-neighbor interactions, since the system with multiple spins extends to $N-1$ pairwise local couplings of this type. 

\begin{figure}[b]
\centering
\caption{
Energy level diagram showing the singlet and triplet states separated by an energy gap determined by the exchange coupling $J$. The relative positioning depends on the sign of $J$, with the singlet state becoming the ground state for $J>0$ and the triplet states becoming degenerate ground states for $J<0$.}
\includegraphics[width=1\textwidth]{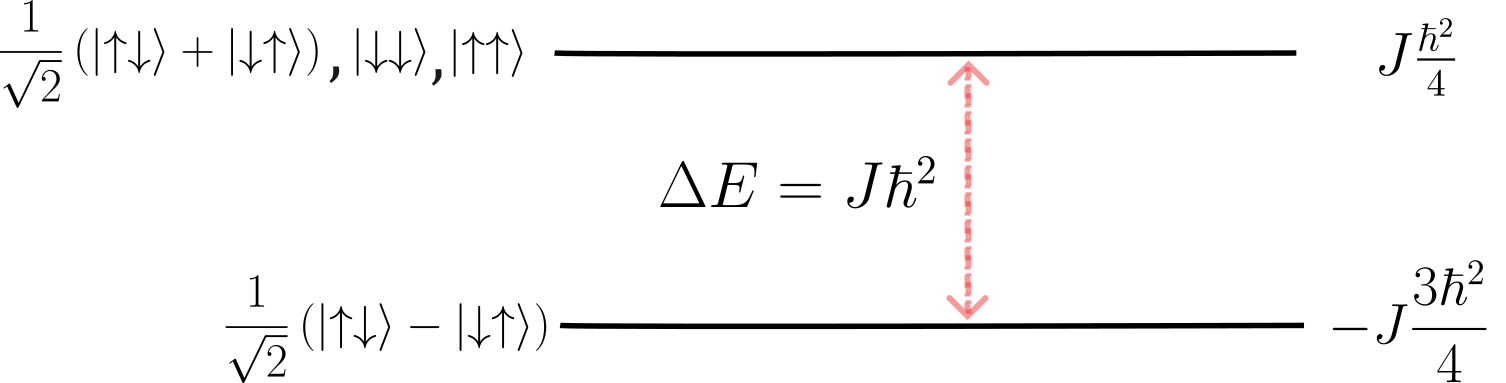}
\caption*{\small Source: The author.}
\label{fig:energy_levels}
\end{figure}

The Pauli Exclusion Principle mandates that the total wavefunction of two electrons must be antisymmetric under particle exchange. This fundamental constraint leads to two distinct cases for the combined spin states. When the spatial wavefunction is antisymmetric, the spin wavefunction must be symmetric, yielding three possible configurations:
\begin{align}
\chi_1(s_1,s_2) &= \ket{\uparrow\uparrow}, \\
\chi_2(s_1,s_2) &= \ket{\downarrow\downarrow}, \\
\chi_3(s_1,s_2) &= \frac{1}{\sqrt{2}}(\ket{\uparrow\downarrow} + \ket{\downarrow\uparrow}).
\end{align}
These symmetric spin states collectively form the triplet configuration with total spin quantum number $s=1$. Direct calculation confirms that application of the total spin operator $\mathbf{S}^2$ yields
\begin{equation}
\mathbf{S}^2\chi_i = 2\hbar^2\chi_i \text{ ,} \quad \text{for all triplet states.}
\end{equation}
Conversely, when the spatial wavefunction is symmetric, the spin component must be antisymmetric, resulting in a single unique state
\begin{equation}
\chi_4(s_1,s_2) = \frac{1}{\sqrt{2}}(\ket{\uparrow\downarrow} - \ket{\downarrow\uparrow}).
\end{equation}
This antisymmetric configuration forms the singlet state with $s=0$, as verified by
\begin{equation}
\mathbf{S}^2\chi_4 = 0.
\end{equation}
This analysis reveals that exchange interactions give rise to precisely two possible configura-\linebreak tions for the system, triplet state or a singlet state. In the lab framework, the system will spontaneously adopt whichever configuration corresponds to the ground state, that is, the energetically favorable state determined by the system parameters.
With this in mind, Heisenberg introduced his model 
\begin{equation}
H_{12} = J \mathbf{S}_1 \cdot \mathbf{S}_2.
\end{equation}
To understand why this is associated with the exchange interaction we look at its energy spectrum. Upon diagonalization we find that there are two energy levels associated with the triplet and singlet states mentioned above. For the triplet states we have,
\begin{equation}
E_{\text{triplet}} = \frac{J\hbar^2}{4},
\end{equation}
while the singlet state occupies a separate energy level,
\begin{equation}
E_{\text{singlet}} = -\frac{3J\hbar^2}{4}.
\end{equation}
The exchange coupling constant $J$ serves as the fundamental parameter governing both the ground state configuration and energy level structure of the system. For $J > 0$, the system energetically favors the singlet state with its antisymmetric spin configuration, while $J < 0$ sets the symmetric triplet states as the ground state. This structure is schematically represented in Fig. \ref{fig:energy_levels}.

The magnitude $|J|$ directly sets the energy scale separating these eigenstates. Stronger coupling produces more substantial energy differences, leading to greater separation between the singlet and triplet states. This tunable energy gap provides precise control over quantum properties, enabling accurate modeling of physical systems through adjustment of this single parameter. Conversely, this simplicity facilitates theoretical analysis, by focusing on variations in $J$, we can efficiently map theory to experiments and \textit{vice-versa}. 

From now on we set $\hbar=1$.
\section{PROTOCOLS 1 AND 2}

\begin{figure}[b]
    \centering
     \caption{
    (a) P1 and (b) P2 architectures. Bold lines represent $\Delta$ couplings, while thin lines indicate $\delta$
couplings.}
\includegraphics[width=0.7\linewidth]{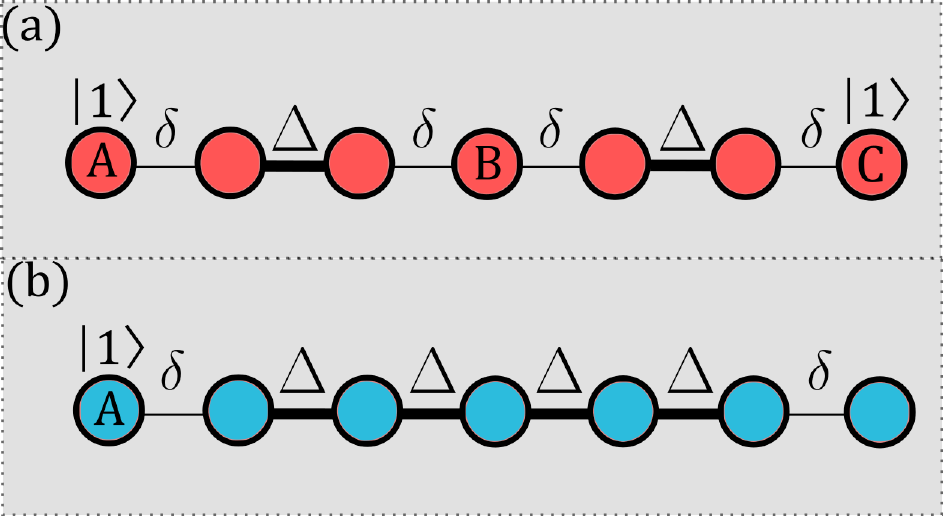}
   \caption*{\small Source: The author.}
    \label{fig:fi1}
\end{figure}

P1 and P2 differ in both the configuration and the initialization
of the spin chains (SCs), as illustrated in Fig.~\ref{fig:fi1}. In both cases, the
dynamics are governed by an XX Hamiltonian,
\begin{align}
    \hat{H} = J \sum_{i=1}^{N-1} \left( \hat{S}_i^x \hat{S}_{i+1}^x + \hat{S}_i^y \hat{S}_{i+1}^y \right)
    + \sum_{i=1}^{N} B_i \hat{S}_i^z,
\end{align}
where the nearest-neighbor coupling \(J\) takes the values \(\Delta\) or \(\delta\)
depending on the local interaction.

Throughout this dissertation, we focus on chains of length \(N = 7\).
This choice is not arbitrary but is imposed by structural constraints intrinsic to P1: due to its dimerized coupling pattern, P1 operates effectively only for
chains with an odd number of sites of the form \(N = 2m + 1\), where a unique central
site mediates the effective trimer dynamics.
By contrast, P2 does not rely on dimerization-induced localization and can,
in principle, be implemented for arbitrary chain lengths.
Restricting the analysis to \(N = 7\) therefore enables a fair and controlled
comparison between the two protocols, while still capturing the essential physical
mechanisms governing their entanglement generation dynamics.

P1 is initializided in the state 
\begin{align}
    |\psi_{P1}(0)\rangle = |1\rangle_A \otimes |0\rangle^{\otimes (N-2)} \otimes |1\rangle_C,
    \label{eq:P1initial}
\end{align}
and evolves unitarily with $B_i=0$. In this configuration, the boundary spins A and C are initially excited, while the intermediate sites are unexcited. The entanglement dynamics, in this case, arise from coherent exchange interactions distributed across the entire chain.

For P2 the chain is initialized in the state 
\begin{align}
    |\psi_{P2}(0)\rangle = |1\rangle_A \otimes |0\rangle^{\otimes (N-1)},
\end{align}
here, a single excitation is localized at the sender (A) site, while all other spins, including the receiver (at the opposite end of the chain), begin in the unexcited state. 

It is important to clarify that for a spin-$s$ system, we define the computational basis states in terms of the eigenstates of the $S_z$ operator.
Specifically, for spin-$1$, the basis states are:

\begin{align}
|0\rangle &\equiv |m = -1\rangle, \\
|1\rangle &\equiv |m = 0\rangle, \\
|2\rangle &\equiv |m = +1\rangle.
\end{align}

For spin-${3}/{2}$, they are:
\begin{align}
|0\rangle &\equiv |m = -\tfrac{3}{2}\rangle, \\
|1\rangle &\equiv |m = -\tfrac{1}{2}\rangle, \\
|2\rangle &\equiv |m = +\tfrac{1}{2}\rangle, \\
|3\rangle &\equiv |m = +\tfrac{3}{2}\rangle.
\end{align}

The protocols for higher spin systems are direct analogs of the spin-1/2 case described above.
In P1, the bulk spins are initialized in the minimal $S_z$ eigenstate, $|m = -s\rangle$, corresponding
to the lowest indexed basis state, while the boundary spins are set to the maximal
eigenstate, $|m = +s\rangle$, corresponding to the highest indexed basis state. In contrast, P2 initializes all spins uniformly in the minimal eigenstate $|m = -s\rangle$, except for the sender (A) site, which is initialized in the maximal eigenstate. A zero magnetic field $B_i = 0$ is applied in the bulk, while carefully engineered, optimized boundary magnetic fields
are applied at the extremities to enhance the coherent buildup of long-range entanglement.

A central advantage of P2 is that the bulk (spins $2$ through $N - 1$) remains largely unexcited during evolution. That is, the intermediate spins undergo only virtual excitation, which avoids a significant population of the bulk and enables the boundary spins to interact effectively as if they were directly coupled. This virtual coupling mechanism reduces the influence of imperfections within the chain, such as diagonal and off-diagonal disorder or local dephasing, thereby supporting the robust generation of entanglement between the sender and receiver.

Although structurally reminiscent of state transfer protocols, the goal here is not to maximize the transfer fidelity but to exploit coherent dynamics for the fast and resilient generation of entanglement.

\section{WHY P1?}
\label{sec:P1}
The reader may wonder why these two distinct configurations are necessary. While they may appear arbitrary at first glance, they are in fact carefully chosen. The initial state of the system is critically important for two reasons: first, it is one of the only two points of interaction of the user with the system, and second, it fundamentally determines the final entangled state that is produced.

The specific distribution of spin-spin interactions is designed to generate a maximally entangled state for a given initial state in the spin-1/2 case. This design principle naturally extends to systems with larger spin dimensions ($s = 1$, $3/2$). However, as the spin dimension increases, the maximum achievable entanglement decreases.

This reduction in entanglement occurs because the higher-dimensional SCs have a larger computational basis, that is, the two end spins have more quantum states available to them. This increased number of available states makes it significantly more difficult to concentrate the system's correlations into a single, maximally entangled pair, which is why perfect maximal entanglement becomes harder to achieve.

\begin{figure}[b]
  \centering
  \caption{Effective spin chain with $\eta$ couplings.
  }
  \includegraphics[width=0.7\columnwidth]{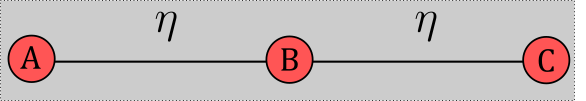}
  \caption*{\small{Source: The author.}}
  \label{fig:fig2}
\end{figure}

P1's configuration can be better understood as  an effective three-site (trimer) model involving
only the sites labeled A, B, and C. For ${\delta}/{\Delta}\ll 1$ the chain effectively breaks into dimers weakly coupled via $\delta$. So, in this regime the dynamics relevant to entanglement generation can be captured by an effective trimer model involving only the boundary spins $A$, $C$ and the central site $B$, as  shown in Fig. \ref{fig:fig2}.

The effective Hamiltonian projected to the single-excitation subspace is 
\begin{equation}
  H_{\mathrm{trimer}} =
  \begin{pmatrix}
    0    & \eta & 0 \\
    \eta & 0    & \eta \\
    0    & \eta & 0
  \end{pmatrix}.
\end{equation}

Diagonalizing this matrix, we obtain 
\begin{align}
|\phi_- \rangle = \frac{1}{2} \begin{pmatrix} -1 \\ \sqrt{2} \\ -1 \end{pmatrix}, \quad 
|\phi_0 \rangle = \frac{1}{\sqrt{2}} \begin{pmatrix} 1 \\ 0 \\ -1 \end{pmatrix}, \quad 
|\phi_+ \rangle = \frac{1}{2} \begin{pmatrix} 1 \\ \sqrt{2} \\ 1 \end{pmatrix},
\end{align}
with \(|\phi_- \rangle\) having energy \(E_- = -\sqrt{2}\eta\), \(|\phi_0 \rangle\) having energy \(E_0 = 0\), and \(|\phi_+ \rangle\) having energy \(E_+ = \sqrt{2}\eta\).

In the case of P1, even though there are two excitations present in the chain, it is not necessary to consider the second-excitation subspace separately. This is due to a particle-hole symmetry present between the one- and two-excitation subspaces, which establishes a direct mapping between the basis states of the one- and two-excitation subspaces:
\begin{align}
\ket{100} \leftrightarrow \ket{011},\quad \ket{010} \leftrightarrow \ket{101}, \quad \ket{001} \leftrightarrow \ket{110}.
\end{align}
The corresponding eigenvalues for the one-excitation and two-excitation subspaces are identical, meaning the energy spectra are the same for both cases. Since the dynamics of the system are governed by the energy eigenvalues, the time evolution in both subspaces is essentially equivalent. 

This symmetry arises because the XX interaction term, $S_x^iS_x^j + S_y^iS_y^j$, is fundamentally a spin-exchange operator. It only acts on the relative alignment of spins, not on their individual orientation. Consequently, its action is invariant under a global spin flip ($\ket{0} \leftrightarrow \ket{1}$ for all sites), which is the origin of the particle-hole symmetry.

This effective model generalizes to longer chains by symmetrically adding dimer pairs
around the central site B, preserving low-energy trimer-like dynamics. However, the effective coupling $\eta$ decreases exponentially with the chain length, leading to a corresponding increase in the entanglement timescale \cite{Estarellas2018}.
The values of the effective coupling $\eta$ can be obtained from the eigenvalues immediately above or below zero in the one-excitation spectrum. When calculating the single excitation spectrum of the 7 sites ABC chain, seven different energy states are obtained. Two on an upper band \( E^{+2} \), \( E^{+3} \), two on a lower band \( E^{-2} \), \( E^{-3} \) and three sitting in between (energy gap), \( E^{+1} \), \( E^0 \) and \( E^{-1} \). From diagonalising the one-excitation subspace of the full Hamiltonian in terms of \(\Delta\) and \(\delta\) we obtain the following analytical forms of such eigenvalues:

\begin{align}
E^{\pm 3} &= \pm \frac{\sqrt{\Delta^2 + 3\delta^2 + \sqrt{\Delta^4 + 6\Delta^2 \delta^2 + \delta^4}}}{\sqrt{2}}, \\
E^{\pm 2} &= \pm \sqrt{\Delta^2 + \delta^2}, \\
E^{\pm 1} &= \pm \frac{\sqrt{\Delta^2 + 3\delta^2 - \sqrt{\Delta^4 + 6\Delta^2 \delta^2 + \delta^4}}}{\sqrt{2}}, \\
E^0 &= 0.
\end{align}
The efficacy of the reduced trimer model is related to two fundamental properties of the original chain's energy spectrum. First, the energy levels $E^{\pm 1}$ and $E^0$ form a low-energy band. This concentration of states near $E^0$ implies that transitions between them require only a small energy exchange. Conversely, transitions from this low-energy subspace to other eigenstates would necessitate a large energy difference. This large energy separation effectively restricts the system's dynamics within this energy band, rendering the remaining states as forbidden for the system's evolution following a proper initial state preparation (e.g. two excitations injected, one on A and other on C).

\begin{figure}[h]
    \centering
    \caption{
    Probability distribution across lattice sites for the eigenstates $E^{-1}$, $E^0$, and $E^{+1}$ (left) and their corresponding energy levels (right) in the 7-site chain with $\Delta/\delta = 10$. The color coding represents different lattice sites: A (dark blue), 2 (orange), 3 (red), B (brown), 5 (pink), 6 (green), and C (light blue). These three energy levels form a subspace that is nearly degenerate and spatially localized on sites A, B, and C, with minimal probability amplitude on the intermediate sites 2, 3, 5, and 6. This spatial localization and energy level approximation validate the effective trimer model approximation, where the system's dynamics are confined to the A-B-C subsystem due to the large energy separation from the upper and lower bands  ($E^{\pm 2}$, $E^{\pm 3}$).}
    \includegraphics[width=\linewidth]{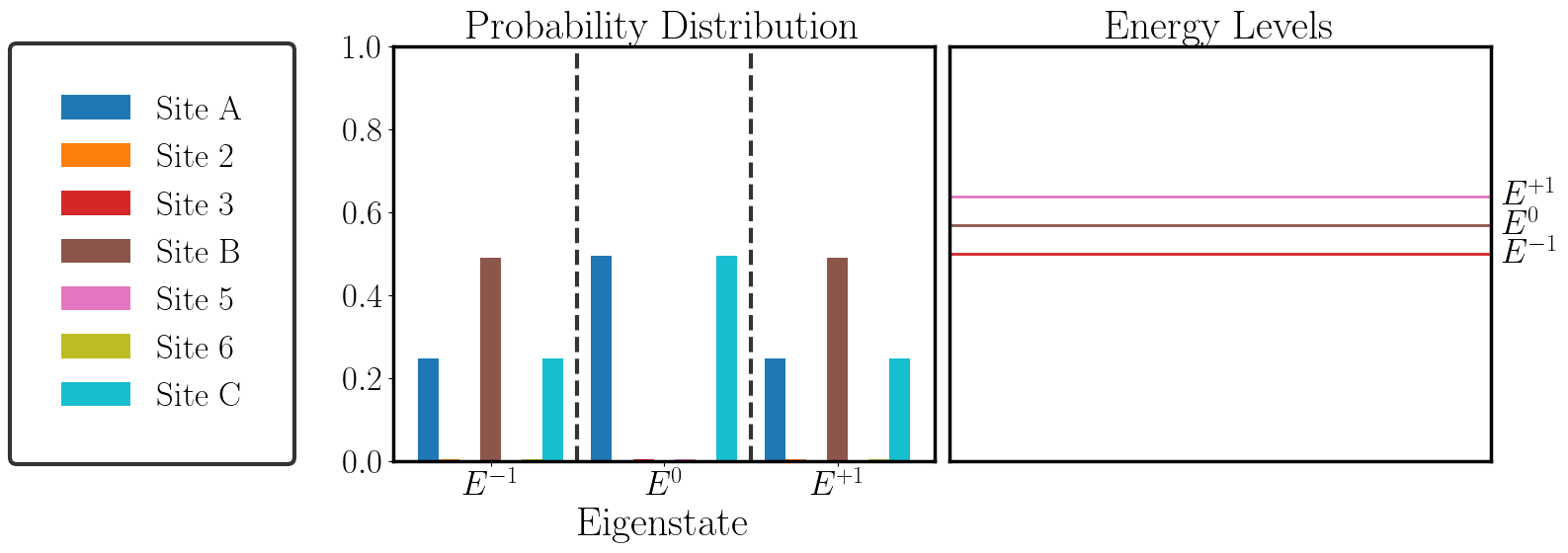}
    \label{fig:probsP1}
    \caption*{\small{Source: The author.}}
\end{figure}

The second crucial factor is the spatial structure of the eigenstates associated with these gap energies. The states corresponding to $E^{\pm 1}$ and $E^0$ are localized on sites A, B, and C, only exhibiting probability of the excitation being found within this sites. Consequently, the system's dynamics are not only confined to a specific energy range but are also spatially constrained to a specific subset of sites. Both of these characteristics, essential for the functioning of P1, are shown in Fig. \ref{fig:probsP1}.

The combination of these two properties ensures that the system's evolution is overwhelmingly governed by the interactions between sites A, B, and C. It is this confinement that causes the complex dynamics of the full chain to emerge naturally from the simple, effective description provided by the three-site trimer model. It is important to note that this only happens for $\delta/\Delta \ll 1$.

This motivates us to equate the eigenvalue $E_+$, found when diagonalizing the trimer chain's Hamiltonian, with $E^{+1}$, found in the diagonalization of the 7-spin chain one-excitation subspace's Hamiltonian,
\begin{align}
E^{+1}=\sqrt{2}\eta &= \frac{\sqrt{{\Delta^2 + 3\delta^2} - \sqrt{\Delta^4 + 6\Delta^2\delta^2 + \delta^4}}}{\sqrt{2}} 
\end{align}
which implies in
\begin{align}
\eta &= \frac{\sqrt{{\Delta^2 + 3\delta^2} - \sqrt{\Delta^4 + 6\Delta^2\delta^2 + \delta^4}}}{{2}}.
\end{align}
This justifies the approximation, and allows to calculate the effective coupling $\eta$. This trimer mediates coherent oscillations between A and C through B, enabling the generation of high-fidelity entanglement. We thus provide an analytic interpretation of the system behavior that demonstrates the importance of the presence of sites A, B and C for the operation of our protocol, mathematically validates the entangled state formation, and gives accurate recipes to obtain a value for $\eta$.

\section{WHY P2?}
\label{sec:P2}
Now, we examine why P2 represents a promising candidate for entanglement generation protocols. Initially, one might consider applying the same approach used for P1. However, numerical analysis of P2's one-excitation subspace Hamiltonian reveals a crucial difference: while the energy level structure shows similar grouping patterns to P1, the eigenstates forming the low-energy subspace are distributed throughout the entire chain rather than being spatially confined. This delocalization prevents the viability of an effective trimer model approximation, as the system dynamics cannot be reduced to interactions between only three specific sites.

Thus, the analysis for this protocol will demand a different method, called Time-Coarse Graining (TCG) \cite{Lee2018,Klimov2002}. It will explore the fact that the two sites coupled in the extremes of P2's chain can be treated as small perturbations in the Hamiltonian of the bulk of the chain. This, in the end, will lead to an effective model, just a different one from the trimer model presented for P1. 

The TCG operation, also called Time-Coarse Grained operator, is defined as
\begin{align}
\overline{O(t)} &\equiv \int_{-\infty}^\infty dt' f(t - t') O(t'), \quad \text{with } \int_{-\infty}^\infty dt' f(t') = 1, 
\label{eq:TCG}
\end{align}
where \( f(t) \) is a low-pass filter function. In particular, we are
interested in the TCG density operator $\overline{\rho(t)}$. The positivity, unit trace, and hermiticity of $\overline{\rho(t)}$ ensure that it remains a valid density operator, just like $\rho(t)$ \cite{Gamel2010}.
By assumption, $\overline{\rho(t)}$ represents the evolution of a closed
system and, therefore, is given by the unitary evolution
\begin{equation}
\overline{\rho(t)} = \hat{U}(t,t_0)\,\overline{\rho(t_0)}\,\hat{U}^\dagger(t,t_0),\label{eq:evo}
\end{equation}
where $\hat{U}(t,t_0)$ is the time-ordered evolution operator,
satisfying the Schr\"odinger equation \cite{sakurai2020modern}
\begin{equation}
i\hbar \frac{\partial}{\partial t}\hat{U}(t,t_0) = \hat{H}(t)\,\hat{U}(t,t_0).
\label{eq:schr}
\end{equation}
To calculate a series expansion for $\hat{U}(t,t_0)$, we adopt the
standard approach of replacing $\hat{H}(t)$ by $\lambda \hat{H}(t)$, where
$\lambda$ is a dimensionless expansion parameter. One can
imagine $\lambda$ gradually increasing from $0$ to $1$, representing
the Hamiltonian being "gradually turned on". When $\lambda=1$ we
recover the standard solution. This suggests that we
can write $\hat{U}(t,t_0)$ as a series in powers of $\lambda$
\begin{equation}
\hat{U}(t,t_0) \equiv \sum_{n=0}^\infty \lambda^n \hat{U}_n(t,t_0). \label{eq:U}
\end{equation}
Substituting into Eq. \eqref{eq:schr} and matching coefficients of similar powers of
$\lambda$, we obtain the recursion relation
\begin{align}
\frac{\partial}{\partial t}\hat{U}_0(t,t_0) &= 0, \\
i\hbar \frac{\partial}{\partial t}\hat{U}_n(t,t_0) &= \hat{H}(t)\,\hat{U}_{n-1}(t,t_0), \quad n\geq 1. \label{eq:recurrence}
\end{align}
Given that for $\lambda \rightarrow 0$ we must have $U(t,t_0)=\mathbbm{I}$ then $U_0(t,t_0)=\mathbbm{I}$.

On substituting Eq. \eqref{eq:U} in Eq. \eqref{eq:evo} one finds an equation of evolution for $\overline{\rho(t)}$ in terms of $U_n$ and $\lambda$. Suppressing, for the moment, the explicit dependence on $t$ and $t_0$, and denoting $\rho_0\equiv \rho(t_0)$, we have
\begin{align}
    \overline{\rho} &= 
\sum_{m,n=0}^{\infty} \lambda^{m+n}\, \hat{U}_m \rho_0 \hat{U}_n^\dagger \\
&= \sum_{k=0}^\infty \lambda^k 
\left( \sum_{j=0}^k \hat{U}_{k-j}\,\rho_0\,\hat{U}_j^\dagger \right) \\
&\equiv \sum_{k=0}^\infty \lambda^k \mathcal{E}_k[\rho_0] 
\;\equiv\; \mathcal{E}[\rho_0],
\end{align}
 where $\mathcal{E}_k[\rho_0] \equiv \left( \sum_{j=0}^k \hat{U}_{k-j}\,\rho_0\,\hat{U}_j^\dagger \right) $ and $\mathcal{E}$ is the linear map that takes from $\rho_0$ to $\overline{\rho}$.

To proceed, we need to find the inverse of this transformation, $\mathcal{F} \equiv \mathcal{E}^{-1}$. That is, 
\begin{align}
    \rho 
&= \mathcal{F}[\overline{\rho}]
= \sum_{k=0}^\infty \lambda^k \mathcal{F}_k[\overline{\rho}].
\end{align}

We will use the fact that when applying $\mathcal{F}$ and $\mathcal{E}$ together we must have the identity transformation (i.e. $\mathcal{F}[\mathcal{E}[\rho]]=\mathbbm{I}\rho$). Thus, postulating that $\mathcal{F}$ may be expanded in powers of $\lambda$ and comparing coefficients for powers of the latter, we find 
\begin{align}
\rho 
&= \mathcal{F}[\mathcal{E}[\rho]] 
= \left( \sum_{m=0}^\infty \lambda^m \mathcal{F}_m \right) 
  \left( \sum_{n=0}^\infty \lambda^n \mathcal{E}_n[\rho] \right) \nonumber \\
&= \sum_{m=0}^\infty \sum_{n=0}^\infty 
   \lambda^{m+n} \, \mathcal{F}_m\!\big[\mathcal{E}_n[\rho]\big]=\lambda^0\mathbbm{I}[\rho]. \label{eq:series}
\end{align}
We can relabel the indices by introducing \( k = m + n \). For each fixed total order \( k \), the index \( m \) ranges from \( 0 \) to \( k \), and accordingly, \( n = k - m = k - j\). This re-indexing transforms the double sum into
\begin{align}
    \rho 
&= \sum_{k=0}^\infty \lambda^k 
   \sum_{j=0}^k \mathcal{F}_j\!\big[\mathcal{E}_{k-j}[\rho]\big] 
= \lambda^0 \mathbbm{I}[\rho].
\end{align}
Comparing coefficients of powers $\lambda$ we find the first few terms in the expansion of $\mathcal{F}$, \begin{align}
\mathcal{F}_{0} &= \mathcal{E}_{0} = \mathbbm{I}, \\
\mathcal{F}_{1} &= -\mathcal{E}_{1}, \\
\mathcal{F}_{2} &= -\mathcal{E}_{2} + \mathcal{E}_{1}[\mathcal{E}_{1}].
\end{align}

Now, we can find the effective evolution equation of $\overline{\rho}$ in terms of $\lambda$. Differentiating Eq. \eqref{eq:series} with respect to time and substituting the relations obtained, one can find 
\begin{align}
i \frac{\partial \overline{\rho(t)}}{\partial t} = i \dot{\mathcal{E}}[\rho(t_0)] = i \dot{\mathcal{E}}[\mathcal{F}[\overline{\rho(t)}]] = \sum_{k=0}^{\infty} \lambda^{k} \left\{ \sum_{j=0}^{k} i\dot{\mathcal{E}}_{j} [ \mathcal{F}_{k-j}[\overline{\rho(t)}] ] \right\}.
\end{align}
We can write this as
\begin{align}
i \frac{\partial \overline{\rho(t)}}{\partial t} = \sum_{k=0}^{\infty} \lambda^{k} \mathcal{L}_{k}[\overline{\rho(t)}].
\end{align}
Evaluating $\mathcal{E}_k$ and $\mathcal{F}_k$, we find, up to second order,
\begin{align}
\mathcal{L}_{0}[\rho] &= i \dot{\mathcal{E}}_{0}[\mathcal{F}_{0}[\rho]] = 0,\\
\mathcal{L}_{1}[\rho] &= i \dot{\mathcal{E}}_{1}[\mathcal{F}_{0}[\rho]] + i \dot{\mathcal{E}}_{0}[\mathcal{F}_{1}[\rho]] = \overline{H}\rho - \rho\overline{H},\\
\mathcal{L}_{2}[\rho] &= \overline{HU_{1}}\rho - \overline{H}\,\overline{U_{1}}\rho + \overline{H\rho U_{1}^{\dagger}} - \overline{H}\rho\overline{U_{1}^{\dagger}} - \rho \overline{U_{1}^{\dagger}H} + \rho\overline{ U_{1}^{\dagger}}\,\overline{H} - \overline{U_{1}\rho H} + \overline{U_{1}}\rho\,\overline{H}.\label{eq:second}
\end{align}
Since we will only expand up to second-order, we explicitate the terms up to $k=2$. With this method, we can effectively understand why P2 is a viable platform for quantum entanglement generation. 

We consider a spin-$1/2$ XX chain of length $N$, subject to a uniform magnetic field and coupled at its ends to two additional qubits (emitter $e$ and receiver $r$), which corresponds to the exact configuration of P2. The total Hamiltonian is written as
\begin{align}
    H_S = H_{\text{bulk}} + H_{\text{er}},
\end{align}
where $H_{\text{bulk}}$ describes the chain itself and $H_{\text{er}}$ accounts for the coupling to the boundary qubits. The bulk dynamics are governed by
\begin{align}
    H_{\text{bulk}} 
    = \sum_{n=1}^{N-1} \Delta \left( S_n^x S_{n+1}^x + S_n^y S_{n+1}^y \right) 
    + \sum_{n=1}^{N} \Omega S_n^z ,
    \label{eq:H_bulk}
\end{align}
with uniform nearest-neighbor coupling strength $\Delta$ and magnetic field $\Omega$.  
After applying the Jordan--Wigner transformation \cite{JordanWigner1928} (see Appendix \ref{apendB}), this becomes
\begin{align}
    H_{\text{bulk}}' 
    &= \sum_{n=1}^{N-1} \Delta \left( c_n^\dagger c_{n+1} + c_{n+1}^\dagger c_n \right)
    + \Omega \sum_{n=1}^N c_n^\dagger c_n ,
\end{align}
where $c_n$ ($c_n^{\dagger}$) is the fermionic annihilation (creation) operator on site $n$ and $\Omega$ denotes the uniform energy shift due to the magnetic field.  

In the single-excitation subspace, we expand a general state as
\begin{align}
    \ket{\phi} = \sum_{n=1}^N \phi_n \ket{n}, 
    \quad \ket{n} = c_n^\dagger \ket{0}.
\end{align}
The eigenvalue problem $H_{\text{bulk}}' \ket{\phi} = E \ket{\phi}$ reduces to the recursion relation
\begin{align}
    -\Delta\big(\phi_{n-1} + \phi_{n+1}\big) + \Omega \phi_n = E \phi_n,
\end{align}
with open boundary conditions $\phi_0 = \phi_{N+1} = 0$.  
This is solved by the ansatz
\begin{align}
    \phi_n^k =\sqrt{\tfrac{2}{N+1}} \sin\!\left(\tfrac{\pi k n}{N+1}\right), 
    \quad k = 1, \dots, N,
\end{align}
yielding eigenvalues
\begin{align}
    E_k = \Omega + 2 \Delta \cos\!\left(\tfrac{\pi k}{N+1}\right).
\end{align}
We can define fermionic operators that diagonalize the Hamiltonian as
\begin{align}
    f_k = \sum_{n=1}^{N} \phi_n^{k*} c_n
        = \sqrt{\tfrac{2}{N+1}} \sum_{n=1}^N 
          \sin\!\left(\tfrac{\pi k n}{N+1}\right) c_n,
\end{align}
which satisfy $\{f_k, f_{k'}^\dagger\} = \delta_{k,k'}$, 
the bulk Hamiltonian diagonalizes as
\begin{align}
    H_{\text{bulk}}' = \sum_{k=1}^N E_k f_k^\dagger f_k.
\end{align}
In P2, two additional qubits are coupled to the ends of the chain via
\begin{align}
    H_{\text{er}} 
    = \Omega \left( S_e^z + S_r^z \right) 
    + \delta \left( S_e^+ S_1^- + S_r^+ S_N^- + \text{H.c.} \right),
\end{align}
where $\delta$ denotes the boundary coupling.  
After JW transformation, this becomes
\begin{align}
    H_{\text{er}}' 
    = \Omega\left( c_e^\dagger c_e + c_r^\dagger c_r \right)
    + \delta  \left( c_e^\dagger c_1 + c_r^\dagger c_N + \text{H.c.} \right).
\end{align}
Collecting all contributions, the effective single-excitation Hamiltonian reads
\begin{align}
    H_S' = \sum_{k=1}^N E_k f_k^\dagger f_k
        + \sum_{k=1}^N \bar{\lambda}_k 
        \left( c_e^\dagger f_k + (-1)^{k-1} c_r^\dagger f_k + \text{H.c.} \right),
\end{align}
where 
\begin{align}
    \bar{\lambda}_k = \delta  \sqrt{\tfrac{2}{N+1}} 
        \sin\!\left( \tfrac{\pi k}{N+1} \right) \label{eq:lambda}
\end{align}
and $N$ now represents the number of sites of the complete (bulk, $e$ and $r$) chain. Transforming to the interaction picture with respect to 
\(
H_0 = \sum_k E_k f_k^\dagger f_k + \frac{\xi}{2}(S_e^z + S_r^z),
\) 
we obtain
\begin{align}
\bar{H}_S(t) = \sum_k \bar{\lambda}_k \Big[ \big( c_e^\dagger f_k + (-1)^{k-1} c_r^\dagger f_k \big) e^{i \zeta_k t} + \text{H.c.} \Big], \label{eq:Hharm}
\end{align}
with detuning \begin{align}
\zeta_k = 2\Omega - E_k. \label{eq:zeta}
\end{align}
The above Hamiltonian belongs to a class of Hamiltonians known as \textit{harmonic time-dependent Hamiltonians}. This class consist of expressions of the type 
\begin{align}
    {H}(t) = {H}_0 + \sum_{k=1}^{N} 
{h}_k \exp(-i\omega_k t) + {h}_k^{\dagger} \exp(i\omega_k t). \label{eq:generalclass}
\end{align}
 Clearly, Eq. \eqref{eq:Hharm} is simply this Hamiltonian with $H_0=0$, $h^{\dagger}_k=\big( c_e^\dagger f_k + (-1)^{k-1} c_r^\dagger f_k \big)$ and $\omega_k=\zeta_k$. So, we apply the TCG transformation to Eq. \eqref{eq:generalclass}, assuming the averaging kernel $f(t)$ is an ideal low-pass filter and that $H_0=0$. For this class of Hamiltonians, a typical approach is to consider that terms that oscillate with $\omega_{k}$ or $\omega_{k}+\omega_{k'}$ are filtered out and that terms $\omega_{k}-\omega_{k'}$ pass the filter only when $\omega_k=\omega_{k'}$, implying that,
\begin{align}
\overline{\exp(\pm i\omega_kt)} &= 0, \\
\overline{\exp[\pm i(\omega_k+\omega_{k'})t]} &= 0, \\
\overline{\exp[\pm i(\omega_k-\omega_{k'})t]} &= \delta_{kk'}.
\end{align}
Thus, only terms oscillating at frequency $\omega_{k} - \omega_{k'}$ where $\omega_{k} = \omega_{k'}$ yield a non-zero contribution. The first step in obtaining the effective Hamiltonian for this specific class is to evaluate ${U}_1$ using the recurrence relation from Eq. \eqref{eq:recurrence}
\begin{align}
U_1(t,t_0) &= \frac{1}{i} \int_{t_0}^{t} H(t_1)\, dt_1 \\
&= \frac{1}{i} \int_{t_0}^{t} \Bigg[\sum_{k=1}^N 
\left( h_k e^{-i\omega_k t_1} + h_k^\dagger e^{i\omega_k t_1} \right) \Bigg] dt_1 \\[6pt]
&=  \frac{1}{i}\sum_{k=1}^N 
\left[
\frac{h_k}{-i\omega_k} e^{-i\omega_k t_1}
+ \frac{h_k^\dagger}{i\omega_k} e^{i\omega_k t_1}
\right]_{t_0}^{t}\equiv [V_1(t)-V_1(t_0)]
\end{align}
 Now, we apply the chosen filter to $H(t)$, $V_1(t)$ and $U_1(t)$, obtaining
 \begin{align}
     \overline{{H}(t)} = 0, \quad
\overline{{V}_1(t)} = 0, \quad
\overline{{U}_1(t)} = 0.
 \end{align}
 This reduces the effective time-evolution equation, making necessary the evaluation only of the following terms on Eq. \eqref{eq:second},
 \begin{align}
     \mathcal{L}_2[\rho] = \overline{HU_1}\rho + \overline{H\rho U_1^\dagger} - \rho \overline{U_1^\dagger H} - \overline{U_1 \rho H}.
 \end{align}
 To evaluate this terms we simply compute the products and apply the rules we established for time filtering. This results in the following effective Hamiltonian 
\begin{align}
\hat{H}_{\text{eff}} 
=\sum_{k} \frac{1}{ \omega_{k}}
\left[ \hat{h}_k^\dagger, \hat{h}_k \right]
\end{align}
accompanied of the time-evolution equation 
\begin{align}
\dot{\rho} = -i[H_\text{eff},\rho].
\end{align}
If we substitute $h^{\dagger}_k=\big( c_e^\dagger f_k + (-1)^{k-1} c_r^\dagger f_k \big)$, $\omega_k=\zeta_k$ and evaluate this commutator we obtain
\begin{align}
[\hat{h}_k^\dagger, \hat{h}_k] 
&= \bar{\lambda}_k^2 
\Big[
(\hat{c}_e^\dagger \hat{f}_k + (-1)^{k-1}\hat{c}_r^\dagger \hat{f}_k),
(\hat{f}_k^\dagger \hat{c}_e + (-1)^{k-1}\hat{f}_k^\dagger \hat{c}_r)
\Big] \\[6pt]
&= \bar{\lambda}_k^2 \Big(
[\hat{c}_e^\dagger \hat{f}_k, \hat{f}_k^\dagger \hat{c}_e]
+ [\hat{c}_r^\dagger \hat{f}_k, \hat{f}_k^\dagger \hat{c}_r] \nonumber \\
&\quad + (-1)^{k-1} [\hat{c}_e^\dagger \hat{f}_k, \hat{f}_k^\dagger \hat{c}_r]
+ (-1)^{k-1} [\hat{c}_r^\dagger \hat{f}_k, \hat{f}_k^\dagger \hat{c}_e]
\Big) \\[6pt]
&= \bar{\lambda}_k^2 \Big(
\hat{c}_e^\dagger \hat{c}_e - \hat{f}_k^\dagger \hat{f}_k
+ \hat{c}_r^\dagger \hat{c}_r - \hat{f}_k^\dagger \hat{f}_k
+ (-1)^{k-1} \hat{c}_e^\dagger \hat{c}_r
+ (-1)^{k-1} \hat{c}_r^\dagger \hat{c}_e
\Big).
\end{align}
Summing over all the modes $k$ gives the final result, with the effective coupling of the form $\bar{\lambda}_k^2 / \zeta_k$, leading to the second-order effective Hamiltonian  
\begin{align}
H_\text{eff} = \sum_k \frac{\bar{\lambda}_k^2}{\zeta_k} \left( c_e^\dag c_r +  c_r^\dag c_e \right).
\label{eq:Heff}
\end{align}
We truncate the expansion at second order since the system is considered in the dispersive regime ($\bar{\lambda}_k/\zeta_k \ll 1$). Physically, this effective Hamiltonian describes a coherent, second-order virtual process in which the boundary qubits interact directly, rather than through the bulk of the chain.

Because of our filtering choice, which retains only a single frequency component, the dynamics of the averaged system are purely unitary and governed by an effective von Neumann equation. By contrast, if a more permissive filter were employed, allowing multiple close frequencies to pass, the averaging procedure would introduce additional terms in the evolution equation, reflecting decoherence effects arising from the discarded high-frequency components \cite{Gamel2010}. The appearance of these additional terms can be understood as a consequence of coarse-graining: by filtering out high-frequency components, the effective description no longer accounts for the full unitary dynamics of the system.

This mediated interaction allows for the generation of coherent entanglement without significant excitation of the channel. This is why P2 is a good candidate for an entanglement generation protocol.

\section{COMPARING THE PERFORMANCES OF P1 AND P2}

With the motivation for selecting both chains as platforms for our entanglement generation protocols established, we now turn to a comparison of their performance. In this section, we analyze how protocols P1 and P2 behave under identical conditions. This comparison will allow us to identify the strengths and limitations of each protocol, providing insight into their suitability for practical implementations.

Our results show that P2 reaches its maximum end-to-end entanglement value significantly earlier than P1 across $s=1/2$, $1$ and $3/2$. Also, for these same spin values, P2 reaches a higher entanglement value than P1. 

\begin{figure}[p] 
\centering
\caption{Time evolution of the end-to-end negativity for \textbf{(a)} $s = 1/2$, 
\textbf{(b)} $s = 1$, and \textbf{(c)} $s = 3/2$. 
All traces correspond to the same dimerization ratio $\Delta/\delta = 10$. 
The red line represents P1, while the blue line represents P2.}
\begin{subfigure}{0.7\textwidth}
\includegraphics[width=\textwidth]{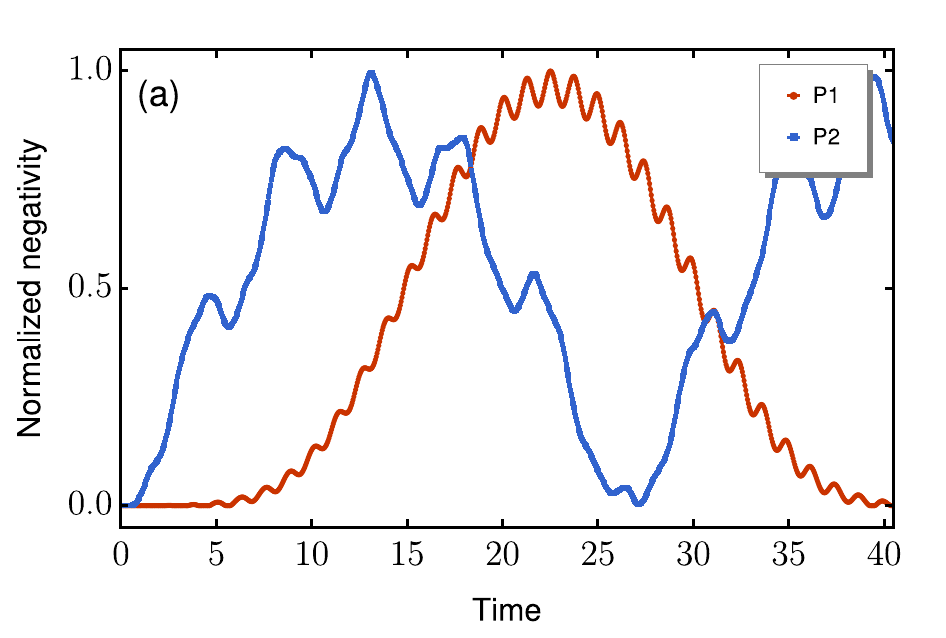}
\end{subfigure}
\begin{subfigure}{0.7\textwidth}
\includegraphics[width=\textwidth]{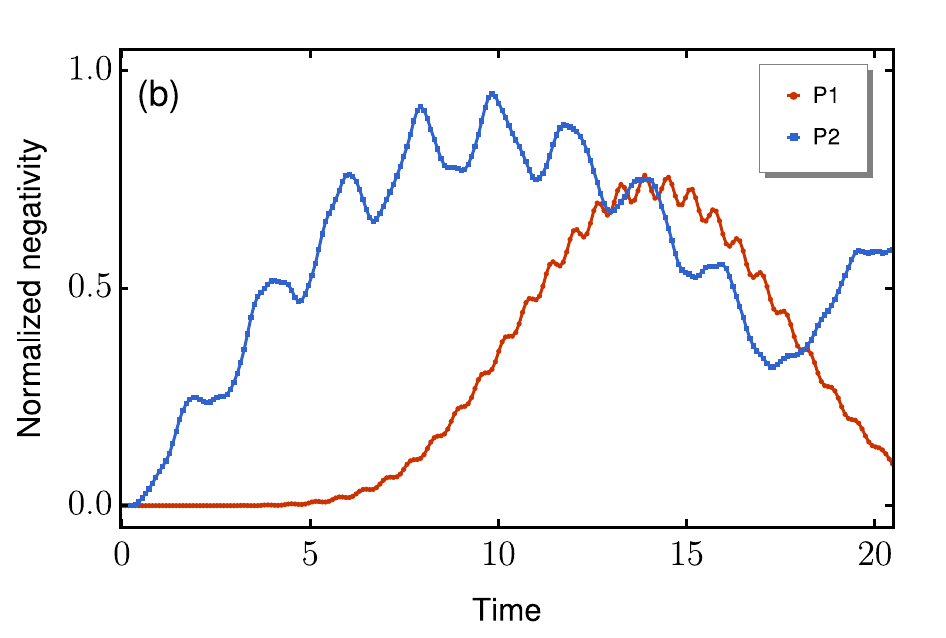}
\end{subfigure}
\begin{subfigure}{0.7\textwidth}
\includegraphics[width=\textwidth]{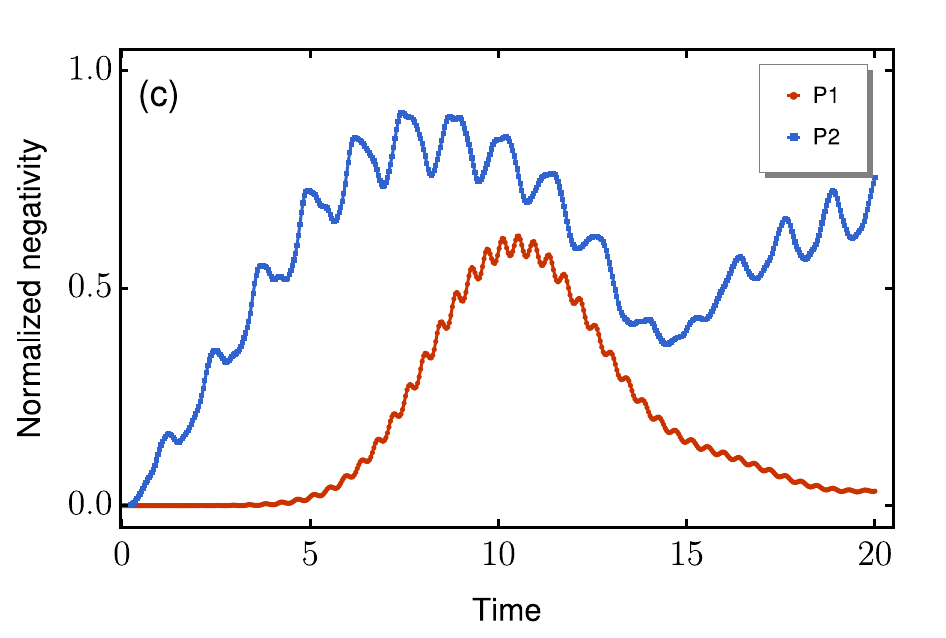}
\end{subfigure}
\caption*{\small{Source: The author.}}
\label{fig:negativityP1P2}
\end{figure}

\begin{figure}[h]
  \centering
  \caption{
    Time evolution of fidelity when considering $\ket{\psi^+}$ as the
    target state. The red line represents P1, while the blue line
    represents P2.
    The time at which maximal entanglement is achieved for each protocol
    is marked with a dashed vertical line.
    We set the dimerization ratio to $\Delta/\delta = 10$.
  }
  \includegraphics[width=0.7\columnwidth]{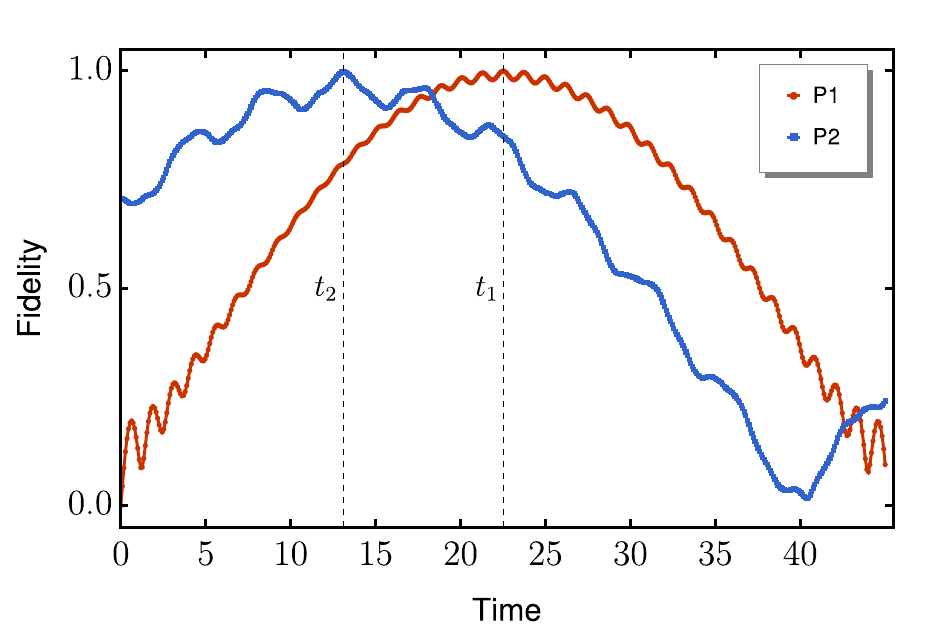}
  \caption*{{\small Source: The author.} }
  \label{fig:fig3}
\end{figure}

A particularly interesting case, due to its technological relevance, is the $s=1/2$ (qubit) system. In modern quantum technologies, qubits are by far the most commonly used, although $s=1$ and $s=3/2$ systems also have practical applications \cite{Schfer2014,vassen2012cold,PhysRevA.71.052318,agarwal2024creatingtwoquditmaximallyentangled}.
In the qubit scenario, both P1 and P2 lead to a maximally entangled state, as shown in Fig.~\ref{fig:negativityP1P2} (a). In fact, both protocols produce the Bell state
\begin{align}
|\psi^+\rangle = \frac{|01\rangle + |10\rangle}{\sqrt{2}},
\end{align}
when the entanglement reaches its maximum, as one can see on Fig. \ref{fig:fig3}, although P2 requires a local \(-\pi/2\) rotation around the \(z\)-axis on the qubit at site \(N\), implemented via the \(R_z\) gate \cite{PhysRevA.84.012301}. It is important to highlight that the state generated by P2 is already maximally entangled, regardless of any local unitary (LU) rotation. The LU rotation applied to P2's state in Fig.~\ref{fig:fig3} serves only as a visualization aid, allowing a direct comparison with the familiar Bell state \(\ket{\psi^+}\). This makes it explicit that both protocols reach a pure, maximally entangled state, even though P2 does so up to a phase.

This highlights an important practical feature of both P1 and P2. In an experimental implementation, once the initial state preparation and the final extraction are performed, no further operations are required. Eliminating additional control steps minimizes the accumulation of errors and enhances the overall robustness and reliability of both protocols.

\clearpage 

The behavior for spin $s=1$ is shown in Fig. \ref{fig:negativityP1P2} (b). In this case, negativity only approaches the maximum possible value for both protocols, yet P2 achieves a higher peak value at an earlier time. This performance trend continues for the $s=3/2$ case in Fig. \ref{fig:negativityP1P2} (c). Again, negativity remains below unity, but P2 reaches a higher value, and does so faster than P1.

\begin{figure}[h]
  \centering
    \caption{
    Contour plot of negativity values as a function of time and the
    magnetic field applied to the boundaries for a $N=7$ chain.
    We set the dimerization ratio to $\Delta/\delta = 10$.
  }
  \includegraphics[width=0.6\columnwidth]{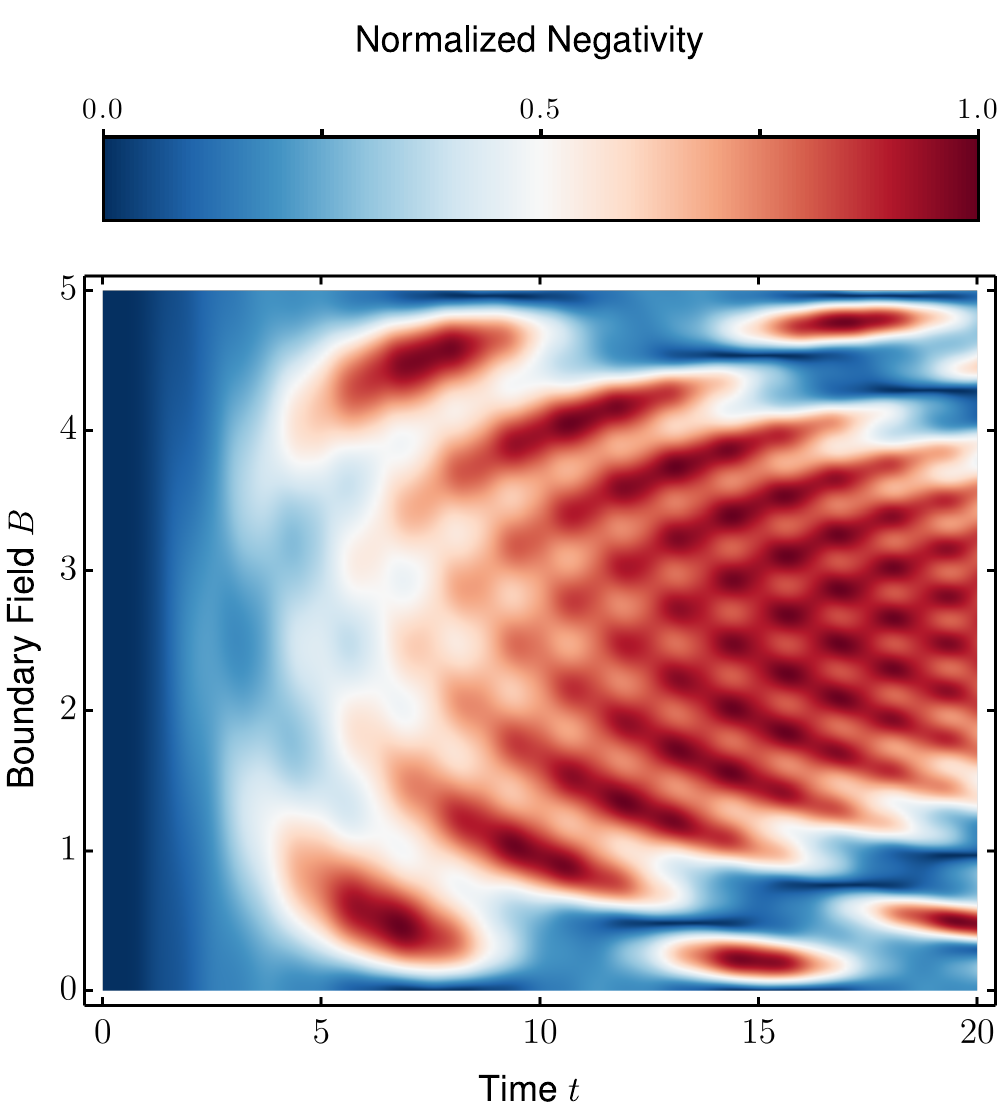}
  \caption*{\small{Source: The author.}}
  \label{fig:fig4}
\end{figure}

Since the performance of P2 depends on the boundary magnetic field \(B\), we numerically determined the value of \(B\) that maximizes the end-to-end negativity for each spin value. 
\begin{table}[b]
\centering
\caption[Peak end-to-end negativity for protocols P1 and P2]{%
Peak end-to-end negativity $\mathcal{N}$ and corresponding evolution
time $t$ for protocols P1 (subscript $1$) and P2 (subscript $2$).
Times are given in units of the weak coupling $\delta$.
The optimal boundary fields are $B_A = B_B = B$.}
\label{tab:negativity_table}
\renewcommand{\arraystretch}{1.3} 
\setlength{\tabcolsep}{12pt}      
\begin{tabular}{cccccc}
\toprule
\textbf{Spin} & $\boldsymbol{\mathcal{N}_{1}}$ & $\boldsymbol{t_{1}/\delta}$ 
              & $\boldsymbol{\mathcal{N}_{2}}$ & $\boldsymbol{t_{2}/\delta}$ 
              & $\boldsymbol{B}$ \\
\midrule
$1/2$ & $1.00$ & $22.50$ & $1.00$ & $13.00$ & $3.7$ \\
$1$   & $0.75$ & $13.90$ & $0.94$ & $9.80$  & $2.9$ \\
$3/2$ & $0.62$ & $10.54$ & $0.90$ & $7.45$  & $4.7$ \\
\bottomrule
\end{tabular}
\end{table}
The optimal parameters extracted from this analysis are listed in Table~\ref{tab:negativity_table}, corresponding to the same curves shown in Fig.~\ref{fig:negativityP1P2}. The full dependence of the negativity on both time and boundary field is displayed in the contour plot of Fig.~\ref{fig:fig4}, which illustrates how the optimal \(B\) is determined from the numerical evaluation.

The analysis in this chapter was focused on systems under idealized conditions. To assess the practical viability of these entanglement generation protocols, however, experimental imperfections must be taken into account. In the next chapter, we address this by introducing and analyzing the effects of static disorder on the performance of both protocols.

\chapter{DISORDER} \label{chap:gipa}

\vspace{0.8cm}
The fabrication of any quantum device is inevitably prone to imperfections because quantum phenomena are highly sensitive to perturbations. Even small deviations can severely compromise the device's intended operation. To account for these effects, we model fabrication-induced errors as random variations in the system's Hamiltonian parameters and evaluate the performance of the entanglement generation protocols across $1000$ realizations, averaging the resulting negativity. We also plot the standard deviation to show how much, on average, the results of each protocol deviate from the mean; this can be interpreted as an indicator of the protocol's experimental reproducibility. Specifically, for each realization, we simulate the time evolution under
the corresponding disordered Hamiltonian, record the maximum
entanglement value attained during the evolution, and then average this
value over all realizations.

The resulting data points, plotted as a function of disorder, quantify the robustness of both protocols against realistic experimental imperfections. Disorder is introduced in both diagonal and off-diagonal terms: the former represents local fabrication defects and on-site field fluctuations, which are among the most detrimental sources of static decoherence in the single-excitation sector of spin chains \cite{Ronke2011,Estarellas2018}, while the latter accounts for fluctuations in the coupling strengths between neighboring sites, modeling imperfections in the interaction between nearest-neighbors that can disrupt coherent transport along the chain.

This practical challenge of maintaining precise control over a quantum system connects directly to the foundational requirements for quantum information processing. Before analyzing our specific disorder models, it is therefore essential to understand the broader criteria that real-world quantum devices and protocols must satisfy.

\section{THE DIVINCENZO CRITERIA}
Our work is motivated by the practical advantages of spin chains, particularly their natural compatibility with solid-state platforms, which makes them promising candidates for quantum technology interfaces. However, this very compatibility introduces fabrication challenges that must be overcome. The essential framework for assessing any quantum platform, including spin-chain-based systems, was established in the seminal paper \cite{Divincenzo2000}, where DiVincenzo outlined seven criteria for the physical realization of quantum information processing. These criteria provide crucial context for understanding why disorder poses such a significant threat to reliable quantum operation.

\begin{tcolorbox}[colback=mygrey, colframe=mygrey, boxrule=0pt, arc=0pt, enhanced jigsaw]
   \textbf{1.}  We Need a Scalable Physical System with Well-Characterized Qubits
\end{tcolorbox}
The first requirement is that the system must consist of a collection of well-characterized qubits. This means accurately knowing all the physical parameters associated with the qubits in the system, which allows us to clearly distinguish the two levels the qubit can occupy. This criterion highlights one of the key challenges in practical quantum computing: scalability. While there is ongoing debate about the exact number of qubits needed to implement "life changing" quantum algorithms (e.g., breaking AES-256 \cite{Grassl2015}), it is universally agreed that a large number of interacting qubits is required. Each additional qubit increases the system's complexity, as more parameters must be precisely controlled and characterized. Fabrication disorder directly threatens this criterion by introducing unpredictable variations in these precisely characterized parameters.

\begin{tcolorbox}[colback=mygrey, colframe=mygrey, boxrule=0pt, arc=0pt, enhanced jigsaw]
   \textbf{2.} There must be a way to control what state is the system initialized to
\end{tcolorbox}

A fundamental requirement for any computational system is the ability to initialize to a known state. This control over the initial conditions is essential for predicting computational outputs and maintaining coherent operations throughout the computation process. In quantum systems, we particularly seek to initialize qubits into low-entropy states, with pure states being optimal as they minimize von Neumann entropy and provide well-defined starting points for quantum operations.

Two primary approaches exist for preparing qubits in a desired state. First, when the target state corresponds to the ground state of the system's Hamiltonian, initialization can be achieved through thermal relaxation, where the system naturally evolves to its lowest-energy configuration through interaction with its environment. Second, projective measurement offers an alternative initialization mechanism, where the quantum state collapses into an eigenstate of the measurement operator, from which the desired state can be prepared through appropriate unitary operations if necessary. Both methods present distinct advantages and challenges in practical implementations, with the choice between them often determined by the specific physical realization of the qubits and the constraints of the experimental setup. The ability to reliably initialize quantum systems remains a critical prerequisite for all subsequent quantum operations and error correction protocols. 

\begin{tcolorbox}[colback=mygrey, colframe=mygrey, boxrule=0pt, arc=0pt, enhanced jigsaw]
   \textbf{3.} The characteristic operation time must be very small when compared to the decoherence time.
\end{tcolorbox}
Decoherence characterizes the dynamics of a qubit in contact with an environment. We can understand it as the time it takes for a pure state to be converted into a mixed state. So, as such disturbances affect the state of the qubits, the integrity of the quantum information processing is compromised at those timescales. So, we require that the computation runs in a much smaller time scale than that necessary for decoherence to affect our system. This criterion will be particularly important when dealing with open quantum systems.
\begin{tcolorbox}[colback=mygrey, colframe=mygrey, boxrule=0pt, arc=0pt, enhanced jigsaw]
   \textbf{4.} There must be a \textit{universal} set of quantum gates.
\end{tcolorbox}
The implementation of quantum algorithms requires the ability to perform arbitrary unitary operations on the system's qubits. In principle, this can be achieved through precise control of the system Hamiltonian that generates the desired unitary transformations. However, experimental constraints severely limit the range of Hamiltonians that can be practically implemented \cite{PhysRevA.51.1015, Barenco_1995}. 

This limitation necessitates the identification of a universal set of quantum gates, a finite collection of operations that can approximate any unitary transformation to arbitrary accuracy through suitable combinations. Crucially, each gate operation must be executed on timescales significantly shorter than the system's decoherence time to maintain quantum coherence throughout the computation. The experimental challenge lies in implementing this universal gate set with both high fidelity and sufficient speed while working within the constraints of available Hamiltonian control. 
\begin{tcolorbox}[colback=mygrey, colframe=mygrey, boxrule=0pt, arc=0pt, enhanced jigsaw]
   \textbf{5.} There must be a way to measure the output of the computation
\end{tcolorbox}
Precise measurement of quantum states represents an essential requirement complementary to state initialization in quantum computation. Consider a qubit described by the density matrix
\begin{align}
\rho = p\ketbra{0} + (1-p)\ketbra{1} + \alpha\ketbra{0}{1} + \alpha^*\ketbra{1}{0}.
\end{align}
An ideal quantum measurement should produce outcome $\ket{0}$ with probability $p = \bra{0}\rho\ket{0}$ and outcome $\ket{1}$ with probability $1-p = \bra{1}\rho\ket{1}$, while simultaneously eliminating quantum coherences through wavefunction collapse. Practical implementations must satisfy two key constraints: (i) the measurement duration must be much smaller than the decoherence time, and (ii) the measurement efficiency must approach as much as possible the ideal case. 

\begin{tcolorbox}[colback=mygrey, colframe=mygrey, boxrule=0pt, arc=0pt, enhanced jigsaw]
   \textbf{6.} There must be a way to convert stationary and flying qubits
\end{tcolorbox}
Stationary qubits are qubits that process quantum information, while flying qubits are responsible for transporting this information.

A quantum computer's hardware will consist mainly of stationary qubits. When long-range computation is needed, flying qubits, typically photons, are used. This makes it necessary to encode the state of static qubits into photonic states.

\begin{tcolorbox}[colback=mygrey, colframe=mygrey, boxrule=0pt, arc=0pt, enhanced jigsaw]
   \textbf{7.} There must be a way to faithfully transmit flying qubits between specified locations
\end{tcolorbox}

The coherence of the flying qubit of choice needs to be preserved until the full state has been transmitted. 

The seven criteria established above provide a rigorous framework for evaluating potential implementations of spin chain-based quantum information processing. Several solid-state platforms have demonstrated promising progress in specific experimental realizations. Among the most notable candidates are: quantum dots \cite{agarwal2023quantum}, superconducting qubits \cite{devoret2004superconductingqubitsshortreview}, trapped ions \cite{Bruzewicz_2019} and NMR-based processors \cite{PhysRevLett.96.170501}. Basically, all these platforms are susceptible to various forms of disorder during fabrication, making the study of disorder effects essential for assessing their practical viability.
\section{DIAGONAL DISORDER}

Against this backdrop of requirements for practical quantum devices, we now examine how fabrication imperfections impact our entanglement generation protocol. Diagonal disorder, also called on-site disorder, tests P2's sensitivity to variations in the fine-tuned boundary magnetic fields, which are essential to its operation. 

To model this type of error, we introduce random local fields $h_i=Ed_i\delta$ with $d_i\in[-0.5,0.5]$ uniformly distributed, where $E$ scales the disorder strength relative to the weak coupling $\delta$.
Essentially, for acessing diagonal disorder, we modifie the Hamiltonian as \begin{equation}
  H \rightarrow H + E\delta(d_1 S_1^z + d_{N-1} S_{N-1}^z).
\end{equation}
For comparison, we apply identical perturbations to P1, establishing a
performance baseline under equivalent conditions.

\begin{figure}[h]
  \centering
    \caption{
    Average peak negativity as a function of diagonal disorder strength
    $E$.
    A red line represents P1, while P2 is shown as a blue line.
    The red and blue bars indicate the standard deviation from the mean
    for each protocol.
    {We set the dimerization ratio to $\Delta/\delta = 10$.}
  }
  \includegraphics[width=0.7\columnwidth]{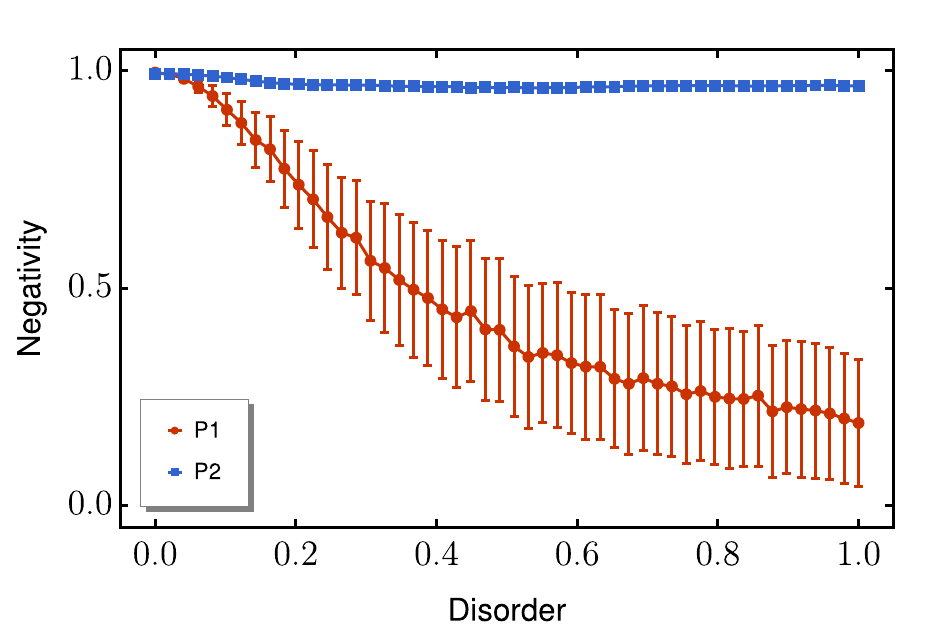}
  \caption*{\small{Source: The author.}}
  \label{fig:fig5}
\end{figure}

Figure \ref{fig:fig5} shows that P2 maintains excellent
performance even at high disorder strengths, while P1 suffers
significant entanglement degradation.
This robustness is particularly valuable for practical implementations,
as it allows for entanglement generation despite imperfections in the applied boundary magnetic fields.
\section{OFF-DIAGONAL DISORDER}

\begin{figure}[h]
  \centering
    \caption{
    Average peak negativity as a function of off-diagonal disorder
    strength $E$.
    A red line represents P1, while P2 is shown as a blue line.
    The red and blue bars indicate the standard deviation from the mean
    for each protocol.
    {We set the dimerization ratio to $\Delta/\delta = 10$.}
  }
  \includegraphics[width=0.7\columnwidth]{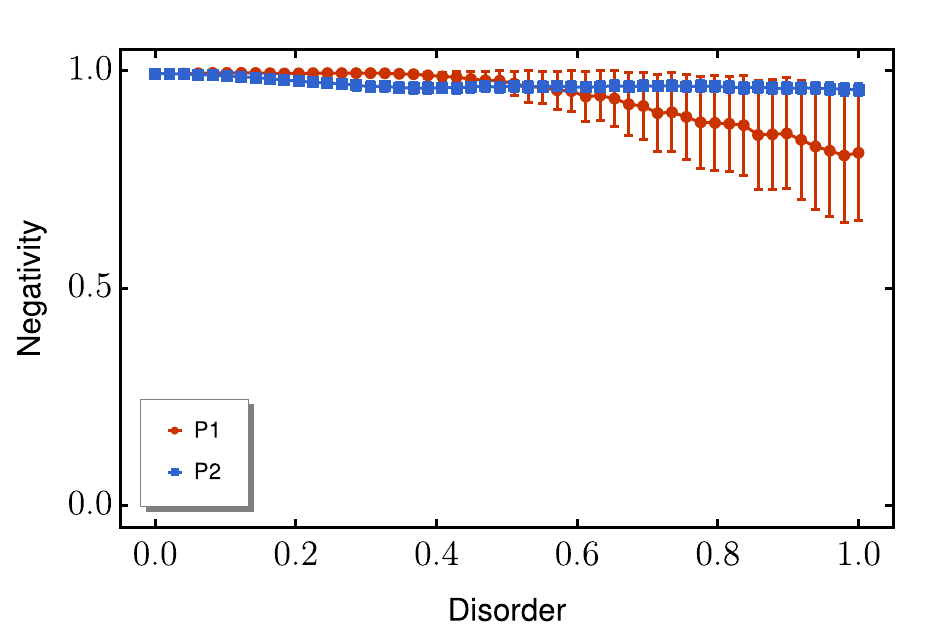}
    \caption*{\small{Source: The author.}}
  \label{fig:fig6}
\end{figure}

\begin{figure}[h]
  \centering
    \caption{
    Average peak negativity as a function of diagonal and off-diagonal disorder
    strength $E$.
    A red line represents P1, while P2 is shown as a blue line.
    The red and blue bars indicate the standard deviation from the mean
    for each protocol.
    {We set the dimerization ratio to $\Delta/\delta = 10$.}
  }
  \includegraphics[width=0.7\columnwidth]{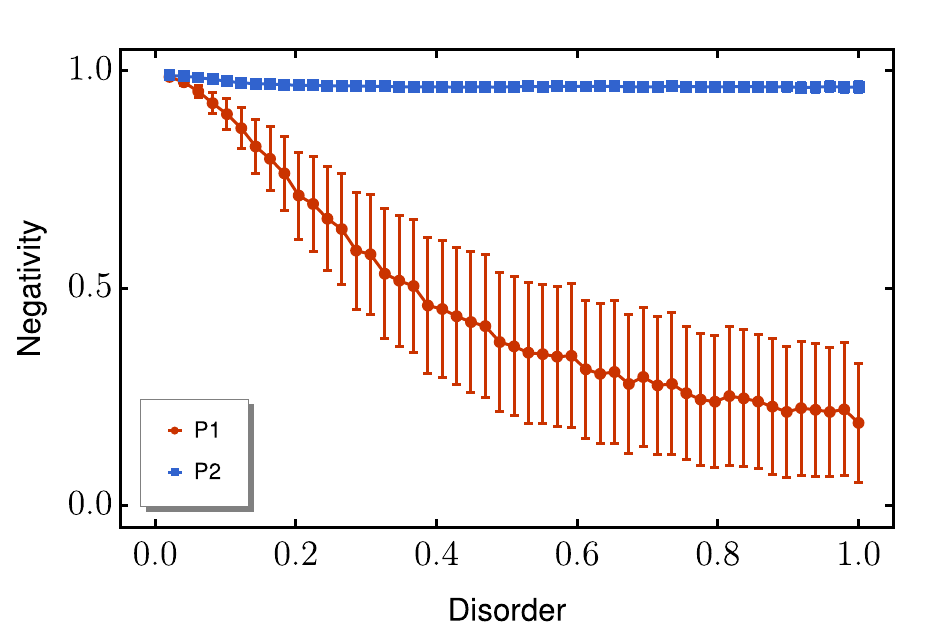}
  \caption*{\small{Source: The author.}}
  \label{fig:fig7}
\end{figure}

The off-diagonal disorder (also called coupling disorder) captures imperfections
in exchange interactions that arise from material defects or control
errors. The modified couplings $J_i \rightarrow J_i + E d_i \delta$
yield the adjusted Hamiltonian:
\begin{equation}
  H \rightarrow
  \sum_{i=1}^{N-1} (J_i + E d_i \delta)
  \left(S_i^x S_{i+1}^x + S_i^yS_{i+1}^y\right).
\end{equation}
For coupling disorder
(Fig. \ref{fig:fig6}), P2 maintains substantial
entanglement ($\mathcal{N} \approx 0.8$) even at values of
$E\approx 0.75\delta$.
Interestingly, when diagonal and off-diagonal disorders are present
simultaneously, P2 continues to outperform P1, as shown in
Fig. \ref{fig:fig7}.
This consistent superiority across all disorder regimes confirms P2's
exceptional resilience to typical solid-state fabrication imperfections.

\clearpage
With this, we assessed the negativity values under different types of disorder. This constitutes an essential step toward approaching the most realistic scenario. However, we can still explore the effects of the environment on the system, particularly dephasing, which can destroy coherence.

\chapter{OPEN QUANTUM SYSTEMS} \label{chap:acoplamentos_dep_temp}
\vspace{0.8cm}

While our discussion has so far focused on closed quantum systems, these represent idealized systems that remain perfectly isolated from their environment. Real physical systems inevitably interact with their surroundings. These interactions may involve energy exchange, information leakage, or other environmental couplings that fundamentally alter the system's dynamics \cite{zhang2025coherent,Landi2022,carrega2016energy,PhysRevA.99.043836}.

Such systems are described by the framework of \textit{open quantum systems}, where we explicitly account for system-environment interactions. Unlike closed systems, the dynamics of open quantum systems generally cannot be described by unitary time evolution alone. Instead, we typically formulate their evolution through equations of motion for the density operator, known as \textit{master equations}. 
These equations are so named because they provide complete knowledge of the system's state, and once solved, the density operator enables calculation of any physical observable's expectation value at any time.

In this chapter, we will primarily focus on {Markovian} (memoryless) dynamics, deriving and analyzing the LME to describe such behavior \cite{Breuer1985, 10.1063/1.5115323}. Additionally, we will briefly examine a method for treating the more complex case of non-Markovian systems.

\section{INTERACTION PICTURE}

Before developing quantum master equations, we must examine the interaction picture. This representation proves particularly valuable when analyzing interactions between a quantum system and its environment. Unlike the Schrödinger or Heisenberg pictures, in which either state vectors or operators alone carry the time dependence, in the interaction picture, both states and operators evolve explicitly in time.

To transition into the interaction picture, we first split the Hamiltonian (written in the Schrödinger picture) into two parts,
\begin{align}
    H = H_S + H_I,
\end{align}
here the subscript $S$ represents the total system's free Hamiltonian and $I$ indicates the interaction Hamiltonian between the system and environment. The state vector in this picture becomes,
\begin{align}
\ket{\psi(t)}_I = e^{iH_St}\ket{\psi(t)},
\end{align}
while a general operator transforms as,
\begin{align}
\hat{A}(t) = e^{iH_St}Ae^{-iH_St},
\end{align}
where the hat identifies operators in the interaction picture and the subscript $I$ designates states in this picture.
Applying this transformation to the VNE yields,
\begin{align}
    \frac{d}{dt}\hat{\rho}(t) = -i[\hat{H}(t),\hat{\rho}(t)].
\end{align}
Often, the VNE in this picture is written in the integral form 
\begin{align}
\hat{\rho}(t) = \hat{\rho}(t_0) - i\int\limits_{t_0}^t ds\, [\hat{H}(s), \hat{\rho}(s)]\,.
\end{align}
This form will be very useful to construct the quantum master equations. 

\section{DYNAMICS OF OPEN QUANTUM SYSTEMS}

In the most general formulation, an open quantum system $S$ constitutes a quantum system coupled to an external quantum system $B$, conventionally designated as a reservoir. The open system thus represents a subsystem of the composite system $S + B$, which, in turn, forms a closed quantum system evolving unitarily under the Schrödinger equation.

\begin{figure}[b]
\centering
\caption{Schematic representation of an open quantum system $S$ interacting with its environment $B$. The dashed boundary emphasizes that $S$ is not isolated, permitting the exchange of energy and information with $B$.}
\includegraphics[width=0.5\linewidth]{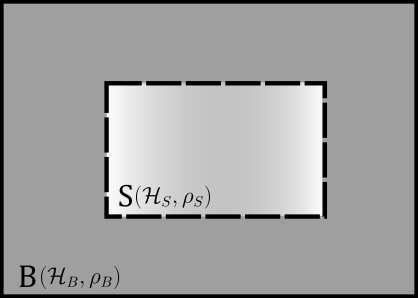}
\caption*{\small{Source: Adapted from \cite{Breuer1985}}.}
\label{fig:env_sys}
\end{figure}

Let $\mathcal{H}_S$ denote the Hilbert space associated with the open system and $\mathcal{H}_B$ the Hilbert space of the environment. The total Hilbert space of the composite system is given by the tensor product space $\mathcal{H} = \mathcal{H}_S \otimes \mathcal{H}_B$. The most general form of the total Hamiltonian $H(t)$ acting on $\mathcal{H}$ admits the decomposition
\begin{align}
H(t) = H_S \otimes \mathbbm{I}_B + \mathbbm{I}_S \otimes H_B + H_I(t),
\end{align}
where $H_S$ represents the free Hamiltonian of the open system $S$, $H_B$ denotes the free Hamiltonian of the environment $B$, and $H_I(t)$ describes the interaction between the system and environment. The operators $\mathbbm{I}_S$ and $\mathbbm{I}_B$ correspond to the identity operators acting on $\mathcal{H}_S$ and $\mathcal{H}_B$ respectively. This decomposition is illustrated schematically in Fig. \ref{fig:env_sys}.

The dynamical evolution of the open quantum system is governed by both its intrinsic dynamics and its interaction with the environment. The system-environment interaction generates correlations between the two subsystems, with the consequence that the reduced dynamics of $S$ can no longer, in general, be described by unitary evolution generated by a Hamiltonian acting solely on $\mathcal{H}_S$. This fundamental characteristic motivates the designation of $S$ as the reduced system, as consideration of $S$ alone necessarily neglects the full dynamical description.

In this work we designate by reservoir, an environment that has an infinite number of degrees of freedom, such that the frequencies of the reservoir modes form a continuous spectrum. This property underlies the emergence of irreversible dynamics in the open system $S$, as energy and quantum information may disperse into the environmental degrees of freedom without recurrence.

A particularly important class of environments are thermal reservoirs, or heat baths, which are defined as environments maintained in a state of thermal equilibrium. The study of open quantum systems coupled to heat baths holds special significance, as such models describe the interaction of quantum systems with their surroundings at finite temperature, a ubiquitous scenario in both natural and engineered quantum systems.

The main motivation of the study of open quantum systems is that in many physically important situations a complete mathematical model of the combined system's dynamics is too complicated. There are multiple ways to contour this and still access the dynamics of the system. Some methods opt by tracing out the environment's and using perturbation theory to obtain an equation of motion for the density operator, this is the case for the Lindblad and Bloch-Redfield master equations \cite{PhysRevA.109.062225}. Methods such as Hierarchical Equations of Motion \cite{doi:10.1143/JPSJ.75.082001} opt by solving a set of coupled differential equations to obtain the dynamics of the system. 

Since the total density matrix $\rho(t)$ evolves unitarily (system $S+B$ is closed) we can write 
\begin{align}
    \rho_{S}(t) = \mathrm{Tr}_{B} \left[ U(t, t_{0}) \rho(t_{0}) U^{\dagger}(t, t_{0}) \right],
\end{align}
where $U(t, t_{0})$ is the time-evolution operator acting on $\mathcal{H}_{SB}$. Since the total system is a pure state, tracing out one of the subsystems will result in a mixed state if the two systems were entangled. Because this is generally the case, we write the VNE for the open system's density matrix, \begin{align}
    \frac{d}{dt}\rho_S(t)=-i\text{Tr}_\text{B}[H(t),\rho(t)].
    \label{eq:von-neumann-system}
\end{align}

\section{THE LINDBLAD MASTER EQUATION}

Now that we have introduced the open quantum system formalism, we can start deriving the equation of motion for the density operator. The standard derivation of the LME rests upon a triad of key assumptions: weak system-environment coupling, the Markovianity of the environment, and the application of the rotating wave approximation. While alternative formulations, such as the Bloch-Redfield equation, can be derived under less restrictive conditions, the advantage of the Lindblad form is its mathematical structure, which guarantees the complete positivity and trace preservation of the density operator. This ensures the solution always remains a physically valid density operator, a crucial property not inherent to all master equations \cite{10.1063/1.5115323}. 

However, this mathematical robustness is achieved at the cost of physical generality, as the required approximations necessarily exclude a range of non-Markovian and strong-\\coupling effects sometimes observed in experimental settings, thereby restricting its scope of application. Since the dynamics generated by the LME is guaranteed to respect the physicality of the state, it is often used in a phenomenological fashion: one guesses
the form of the jump operators on the basis of the kind of processes that the environment is expected to induce, while the rates are parameters to be fitted to experiments \cite{stefanini2025lindbladme}. In this work, we derive the LME from microscopic dynamics, but employ it in a phenomenological manner.

 While the Lindblad approach relies on the idealization of memoryless reservoirs, many realistic environments exhibit finite correlation times, leading to non-Markovian dynamics. In such cases, memory effects can profoundly alter decoherence and entanglement dynamics. A variety of approaches exist to capture these effects, ranging from time-convolutionless master equations to pseudomode mappings, where structured reservoirs are effectively represented by auxiliary damped modes coupled to the system. In this dissertation, we adopt the pseudomode formalism to study non-Markovianity. These methods allow  to retain essential memory effects while still working within a master-equation framework, thereby bridging the gap between the mathematical robustness of the Lindblad form and the richer physical behavior of non-Markovian dynamics.
\subsection{Deduction}

We will consider, without loss of generality, the interaction term as a sum of operators acting simultaneously on the system and environment, 
\begin{align}
    H_I=\sum_i S_i \otimes B_i, \label{eq:Hi}
\end{align}
where $S_i$ belong to the space of operators acting on the system and $B_i$ belong to the space of operators acting on the environment. This decomposition is possible because the set of operators on a finite-dimensional Hilbert space forms a vector space, so any operator on this space can be written as a linear combination of basis operators acting separately on the system and the environment \cite{nielsen_chuang_2010}. Now we utilize equation Eq. \eqref{eq:von-neumann-system} in its integral form, 
\begin{align}
    {\hat{\rho}}(t)=\hat{\rho}(0)-i\alpha\int_0^tds[\hat{H}_I(s),\hat{\rho}(s)].
\end{align}
 This equation can, sometimes, lead to an exact solution but in general this integral is very complicated to solve. The preferred approach to avoid this is to do a perturbative expansion. That is why we introduced a parameter $\alpha$ directly proportional to the interaction strength in the above expression, so we can keep track of what perturbative order each term corresponds to. To do that we substitute the integral form of the VNE into the differential form to obtain \begin{align}
\frac{d\hat{\rho}(t)}{dt} = -i\alpha[\hat{H}_I(t), \hat{\rho}(0)] - \alpha^2 \int_0^t ds\, [\hat{H}_I(t), [\hat{H}_I(s), \hat{\rho}(s)]].
\end{align}
We perform the same procedure once again, 
\begin{align}
    \frac{d\hat{\rho}(t)}{dt} = -i\alpha [\hat{H}_I(t), \hat{\rho}(0)] 
- \alpha^2 \int_0^t ds\, [\hat{H}_I(t), [\hat{H}_I(s), \hat{\rho}(t)]] + \mathcal{O}(\alpha^3). \label{eq:perturbativeser}
\end{align}
At this moment, we perform our first approximation by considering the strength of the interaction between the system and the environment to be small. Therefore, we can avoid higher order terms that wouldn't contribute significantly to the dynamics of the density operator. Now, we have, 
\begin{align}
    \frac{d\hat{\rho}(t)}{dt} = -i\alpha [\hat{H}_I(t), \hat{\rho}(0)] 
- \alpha^2 \int_0^t ds\, [\hat{H}_I(t), [\hat{H}_I(s), \hat{\rho}(t)]]
\end{align}
and trace out the environment's degrees of freedom, \begin{align}
    \frac{d\hat{\rho_S}(t)}{dt} = \mathrm{Tr}_B \left[ \frac{d\hat{\rho}(t)}{dt} \right]
= -i\alpha \mathrm{Tr}_B \left[ \hat{H}_I(t), \hat{\rho}(0) \right]
- \alpha^2 \int_0^t ds\, \mathrm{Tr}_B \left[ \hat{H}_I(t), \left[ \hat{H}_I(s), \hat{\rho}(t) \right] \right].
\end{align}
This is still a very complicated equation to solve because $\hat{\rho}_S(t)$ still depends on the $S+B$ density matrix. To proceed, we assume that at $t=0$ the system and environment are separable, i.e. $\rho(0)=\rho_S(0)\otimes\rho_B(0)$. Physically this corresponds to consider that at $t=0$ system and environment have not interacted yet (or if they did the interactions were short lived). We also assume that the environment corresponds to a thermal reservoir, meaning that it is described by the thermal state \begin{align}
    \rho_B(0)=\frac{\exp{(-H_B/T)}}{\text{Tr}(\exp{(-H_B/T)})}
\end{align}
where $T$ is the temperature and we set $k_b=1$. Using the above assumptions and substituting Eq. \eqref{eq:Hi} we obtain 
\begin{align}
    \mathrm{Tr}_B \left[ \hat{H}_I(t), \hat{\rho}(0) \right]
= \sum_i \left( \hat{S}_i(t) \hat{\rho}_S(0) \, \mathrm{Tr}_B \left[ \hat{B}_i(t) \hat{\rho}_B(0) \right]
- \hat{\rho}_S(0) \hat{S}_i(t) \, \mathrm{Tr}_B \left[ \hat{\rho}_B(0) \hat{B}_i(t) \right] \right).
\end{align}
Now, to calculate the explicit value of this term we assume that $\langle B_i\rangle=\text{Tr}[B_i\rho_B(0)]=0$. This can be done without loss of generality because, supposing that  $\text{Tr}[B_i\rho_B(0)]\neq0$ we can always shift $B_i$ by a constant value $\langle B_i\rangle$, writing $B'_i\equiv B_i-\langle B_i\rangle$. This change only shifts the energy levels, thus not changing the physical action of $B_i$ and now yields $\langle B_i \rangle=0$. Since there is no physical difference between considering $\langle B_i \rangle=0$ or renormalizing it, we assume that we are dealing with operators that satisfy the first, to simplify the calculations. Now, it is clear that,
\begin{align}
    \mathrm{Tr}_B \left[ \hat{H}_I(t), \hat{\rho}(0) \right] = 0,
\end{align}
and the equation of motion reduces to 
\begin{align}
     \frac{d\hat{\rho_S}(t)}{dt}
=
- \alpha^2 \int_0^t ds\, \mathrm{Tr}_B \left[ \hat{H}_I(t), \left[ \hat{H}_I(s), \hat{\rho}(t) \right] \right].
\end{align}
We still need to unravel the system from the environment because this complicates a lot the calculation of the density operator dynamics. We now suppose that system and environment are uncorrelated during all time evolution and any correlation that may appear dies out fast when compared to the system dynamics ($\tau_{corr}<<\tau_{sys}$). Of course this is an approximation, correlations are always expected to appear and they can last for time scales comparable to the system's time scale in some cases. So, when performing this step we are ignoring important effects that would happen experimentally. 
\begin{figure}[h]
    \centering
    \caption{Schematic representation of the system and environmental correlation time scales. The master equation is insensitive to effects occurring on the scale of $\tau_{corr}$. In other words, any phenomena taking place within this time scale will be effectively ignored by the LME.}
    \includegraphics[width=0.7\linewidth]{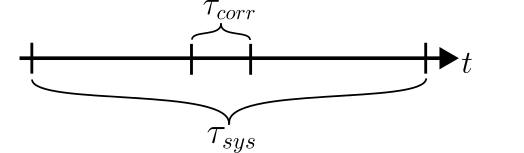}
    \caption*{\small{Source: The author.}}
    \label{fig:tcorrtsys}
\end{figure}

Under this strong assumption we can substitute $\hat{\rho}(t)=\hat{\rho}_S(t) \otimes \hat{\rho}_B(0)$ in the equation of motion to obtain \begin{align}
    \frac{d\hat{\rho_S}(t)}{dt}
=
- \alpha^2 \int_0^t ds\, \mathrm{Tr}_B \left[ \hat{H}_I(t), \left[ \hat{H}_I(s), \hat{\rho}_S(t) \otimes \hat{\rho}_B(0) \right] \right].
\end{align} 
All the terms in the integrand are multiplied by bath correlation terms, $\langle B_i(t) B_j(s) \rangle = \mathrm{Tr}_B \left( B_i(t) B_j(s) \rho_B(0) \right)$, so, because we made the consideration that $\tau_{cor}<<\tau_{sys}$ we can simply extend the upper limit of this integral to infinity with no change to the outcome. Furthermore, since we are considering a stationary bath (i.e. a bath where $\rho_B(t)=\rho_B(0)$), the correlation functions are time-translationally invariant, so they depend only on the difference of time arguments and we can perform the substitution $s \rightarrow t-s$, obtaining,
\begin{align}
    \frac{d\hat{\rho_S}(t)}{dt}
=
- \alpha^2 \int_0^\infty ds\, \mathrm{Tr}_B \left[ \hat{H}_I(t), \left[ \hat{H}_I(s-t), \hat{\rho}_S(t) \otimes \hat{\rho}_B(0) \right] \right].\label{eq:redfield}
\end{align} 

Now, we define the superoperator $\tilde{H}_SA\equiv[H_S,A].$ This is necessary to perform the next approximation we intend to apply. The operator 
\begin{align}
S_i(\omega)=\sum_{\omega = n-m}\bra{n}S_i\ket{m}\ket{n}\bra{m},
\end{align} is an eigenoperator of $\tilde{H}$ and the frequencies $\omega$ correspond to the transition frequencies between the system’s eigenstates that can be induced by the bath. It relates to the original system operator by 
\begin{align}
    S_i=\sum_\omega S_i(\omega).
\end{align}To prove this relation we apply $\tilde{H}$ to $S_i(\omega)$,
\begin{align}
    \tilde{H}S_i(\omega)&=HS_i(\omega) -S_i(\omega)H\\
    &=\sum_{\omega = \epsilon_m-\epsilon_n}H\ket{n}\bra{m}\bra{n}S_i\ket{m}-\sum_{\omega = \epsilon_m-\epsilon_n}\ket{n}\bra{m}H\bra{n}S_i\ket{m}\\
    &=\sum_{\omega =\epsilon_m-\epsilon_n}(\epsilon_n-\epsilon_m)\ket{n}\bra{m}\bra{n}S_i\ket{m}\\
    &=-\omega\sum_{\omega = \epsilon_m-\epsilon_n}\ket{n}\bra{m}\bra{n}S_i\ket{m}\\
    &=-\omega S_i(\omega). \label{eq:eigenoperator}
\end{align}
Taking the Hermitian conjugate on both sides yields 
\begin{align}
    \tilde{H}S^\dagger_i(\omega)&=\omega S^\dagger_i(\omega).
\end{align}
From the above relations, we can see that $S_i(\omega)$ has several convenient properties. First, they obey the eigenvalue equation Eq. \eqref{eq:eigenoperator}, which guarantees that $S(\omega)$ evolves with a simple phase under $H_S$. Second, if $S_i$ is Hermitian, $S_i^\dagger(\omega)=S_i(-\omega)$. These properties indicate a behavior similar to annihilation and creation operators, in the sense that their application on a state reduces or increases its energy by a value $\omega$. 

Now, we will use the eigenoperator spectrum to rewrite the interaction Hamiltonian acting on the system's Hilbert space. First we write this operator in the interaction picture 
\begin{align}
    \hat{H}_I(t)= \sum_i\hat{S}_i(t)\otimes \hat{B}_i(t)
\end{align}
with,
\begin{align}
    \hat{S}_i(t)=e^{iH_St}S_ie^{-iH_St}=\sum_\omega e^{iH_St}S_i(\omega)e^{-iH_St}.
\end{align}
Using the Baker-Campbell-Haussdorf formula \cite{Hall2003BCH}, 
\begin{align}
    e^{iH_St}S_i(\omega)e^{-iH_St}&=S_i(\omega)+it(-\omega S_i(\omega))+\frac{(it)^2}{2!}[H,[H,S_i(\omega)]]+\dots \\
    &=S_i(\omega)+it(-\omega S_i(\omega))+\frac{(it)^2}{2!}(-\omega)^2S_i(\omega)+\dots \\
    &=[1+it(-\omega )+\frac{(it)^2}{2!}(-\omega)^2+\dots]S_i(\omega) \\
    &=e^{-i\omega t}S_i(\omega),
\end{align}
we arrive at 
\begin{align}
    \hat{H}_I(t)=\sum_i \left(\sum_\omega e^{-i\omega t}S_i(\omega)\right) \otimes \hat{B}_i(t)= \sum_{i,\omega} e^{-i\omega t}S_i(\omega) \otimes \hat{B}_i(t) = \sum_{i,\omega} e^{i\omega t}S^\dagger_i(\omega) \otimes \hat{B}^\dagger_i(t),
\end{align}
since $H_I(t)=H^\dagger_I(t)$.

To use this decomposition on Eq. \eqref{eq:redfield} we first expand the commutators
\begin{align}
   \dot{\hat{\rho}}_S(t) = -\alpha^2 \text{Tr}_B \Bigg[ 
\int_0^\infty ds \, \hat{H}_I(t) \hat{H}_I(t - s) \hat{\rho}_S(t) \otimes \hat{\rho}_B(0) \nonumber \\ 
- \int_0^\infty ds \, \hat{H}_I(t) \hat{\rho}_S(t) \otimes \hat{\rho}_B(0) \hat{H}_I(t - s) \nonumber \\ 
- \int_0^\infty ds \, \hat{H}_I(t - s) \hat{\rho}_S(t) \otimes \hat{\rho}_B(0) \hat{H}_I(t) \nonumber \\ 
+ \int_0^\infty ds \, \hat{\rho}_S(t) \otimes \hat{\rho}_B(0) \hat{H}_I(t - s) \hat{H}_I(t) 
\Bigg], 
\end{align}
and from now on we set $\alpha=1$. Now, we apply the eigenvalue decomposition, expand the commutators and utilize the Fourier transformation of the bath correlation functions to absorb the bath contribution to 

\begin{align}
    \Gamma_{kl}(\omega)\equiv \int_0^\infty ds e^{i\omega s}\text{Tr}_B[\hat{B}^\dagger_k(t)\hat{B}_l(t-s)\rho_B(0)],
\end{align}
and obtain 
\begin{align}
\dot{\hat{\rho}}_S(t) = \sum_{\omega, \omega', k, l} 
\bigg( 
e^{i(\omega' - \omega)t} \Gamma_{kl}(\omega) 
\left( 
S_l(\omega) \hat{\rho}_S(t) S_k^\dagger(\omega') 
- 
S_k^\dagger(\omega') S_l(\omega) \hat{\rho}_S(t) 
\right) \\
+ 
e^{i(\omega - \omega')t} \Gamma_{lk}^*(\omega') 
\left( 
S_l(\omega) \hat{\rho}_S(t) S_k^\dagger(\omega') 
- 
\hat{\rho}_S(t) S_k^\dagger(\omega') S_l(\omega) 
\right).
\end{align}
Writing the above expression as commutators,
\begin{align}
    \dot{\hat{\rho}}_S(t) = \sum_{\omega, \omega', k, l} 
\left( 
e^{i(\omega' - \omega)t} \Gamma_{kl}(\omega) 
\left[ S_l(\omega) \hat{\rho}_S(t), S_k^\dagger(\omega') \right] 
+ 
e^{i(\omega - \omega')t} \Gamma_{lk}^*(\omega') 
\left[ S_l(\omega), \hat{\rho}_S(t) S_k^\dagger(\omega') \right]
\right).
\end{align}
At this point, what prevents us from obtaining the standard LME are the terms that explicitly oscillate in time with factors $e^{i(\omega' - \omega)t}$ for $\omega' \neq \omega$. To deal with this terms we will apply the rotating wave approximation, which consists in neglecting these oscillating terms. Physically, this amounts to neglecting processes in which the bath induces  transitions with a large frequency mismatch $\Delta \omega \gg \tau_{sys}^{-1}$. In this approximation, we assume that only the resonant terms with $\omega = \omega'$ contribute significantly to the dynamics. The mathematical justification for this step is that, if the frequency of these oscillating terms is much faster than the system’s dissipative dynamics, they effectively average to zero on the timescales of interest. Therefore, such terms can be neglected,
\begin{align}
    \dot{\hat{\rho}}_S(t) = \sum_{\omega, k, l} 
\left( 
 \Gamma_{kl}(\omega) 
\left[ S_l(\omega) \hat{\rho}_S(t), S_k^\dagger(\omega) \right] 
+ 
 \Gamma_{lk}^*(\omega) 
\left[ S_l(\omega), \hat{\rho}_S(t) S_k^\dagger(\omega) \right]
\right).
\end{align}

It is important to note that, in order to arrive at the LME, neglecting these terms correspond to discarding actual physical processes that may occur in the system. If the full dynamical equation were retained, including these oscillating terms would provide a richer and more accurate description of the system's dynamics at the price of not always mapping to a valid density operator. 

Now, we want to separate the terms that contribute to Hamiltonian dynamics from the non-Hamiltonian ones. For this we will rewrite $\Gamma_{kl}$ as 
\begin{align}
    \Gamma_{kl}=\frac{1}{2}\gamma_{kl}+i\pi_{kl},  
\end{align}
where, 
\begin{align}
\gamma_{kl}=\Gamma_{kl}+\Gamma^*_{kl}=\int_{-\infty}^\infty ds e^{i\omega s}\text{Tr}_B[\hat{B}^\dagger_k(s){B}_l\rho_B(0)],\hspace{1cm} \pi_{kl}=\frac{-i}{2}(\Gamma_{kl}-\Gamma^*_{kl}).
\end{align}
Plugging this definition in the equation of motion we obtain 
\begin{align}
    \dot{\hat{\rho}}_S(t) = \sum_{\omega, k, l} 
\left( 
 \left(\frac{1}{2}\gamma_{kl}+i\pi_{kl}\right)
\left[ S_l(\omega) \hat{\rho}_S(t), S_k^\dagger(\omega) \right] 
+ 
 \left(\frac{1}{2}\gamma_{kl}-i\pi_{kl}\right) 
\left[ S_l(\omega), \hat{\rho}_S(t) S_k^\dagger(\omega) \right]
\right).
\end{align}
Now, we can separate the equation into Hermitian and non-Hermitian parts and transform back to the Schrödinger picture,
\begin{align}
\dot{\rho}_S(t) &= -i[H + H_{LS}, \rho_S(t)] + \sum_{\substack{\omega \\ k,l}} \gamma_{kl}(\omega) \left( S_l(\omega) \rho_S(t) S_k^\dagger(\omega) - \frac{1}{2} \left\{ S_k^\dagger(\omega) S_l(\omega), \rho_S(t) \right\} \right).
\end{align}
where $H_{LS}=\sum_{\omega,k,l}\pi_{kl}(\omega)S^\dagger_k(\omega)S_l(\omega)$. This is usually called a Lamb shift Hamiltonian. In this case, this Hamiltonian only shifts the eigenstate spectrum of the system, as it commutes with the original system Hamiltonian.

The matrix formed by the coefficients \(\gamma_{kl}(\omega)\) is positive as they are the Fourier's transform of a positive function \(\mathrm{Tr}[\hat{E}_{k}^{\dagger}(s)E_{l}\hat{\rho}_{E}(0)]\) \cite{10.1063/1.5115323}. Therefore, this matrix can be diagonalized. This means that we can find a unitary operator \(O\), such that, 
\begin{align}
O\gamma(\omega)O^{\dagger}=\begin{pmatrix}d_{1}(\omega)&0&\cdots&0\\ 0&d_{2}(\omega)&\cdots&0\\ \vdots&\vdots&\ddots&0\\ 0&0&\cdots&d_{N}(\omega)\end{pmatrix}.
\end{align}
We can now write the master equation in a diagonal form,
\begin{align}
\dot{\rho}_S(t)=-i[H+H_{LS},\rho(t)]
+\sum_{l,\omega}\biggl{(}L_{l}(\omega)\rho_S(t)L_{l}^{\dagger}(\omega )-\frac{1}{2}\left\{L_{l}^{\dagger}L_{l}(\omega),\rho_S(t)\right\}\biggr{)}\equiv \mathcal{L}\rho_S(t).
\label{eq:lme}
\end{align}
This is the standard form of the LME. The derivation presented in this section is based on \cite{stefanini2025lindbladme, Breuer1985, 10.1063/1.5115323, Gorini1976, lindblad1976generators}.
\section{RESISTANCE TO DEPHASING}

\begin{figure}[h]
  \centering
    \caption{
    Peak end-to-end negativity as a function of the boundary dephasing
    rate $\gamma$, shown for two coupling regimes: $\Delta = 10$ and
    $\Delta = 30$, with $\delta = 1$.
    The main plot compares the performance of protocols P1 and P2,
    highlighting the enhanced robustness of P2, which exhibits a slower
    decay of entanglement under increasing dephasing.
    The inset displays the maximum population in the bulk channel for
    each protocol, demonstrating that P2 maintains significantly lower
    excitation in the intermediate spins across both coupling regimes.
  }
  \includegraphics[width=0.75\linewidth]{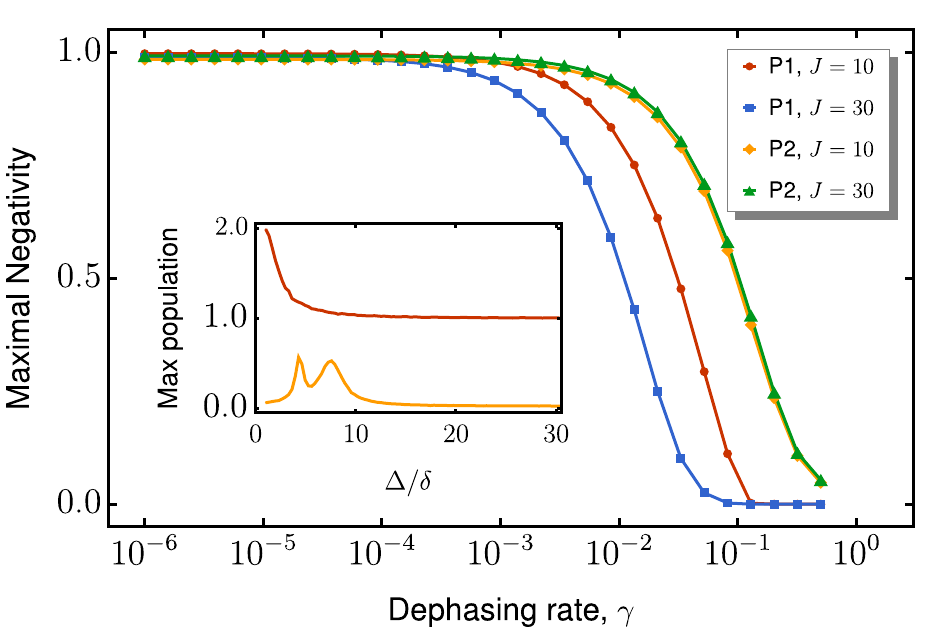}
  \caption*{\small{Source: The author.}}
  \label{fig:fig8}
\end{figure}

A ubiquitous source of decoherence in solid-state devices is the pure dephasing of the qubits that terminate the spin chain and interface it with external control circuitry. Thus, it is valuable to examine how P1 and P2 behave under these effects. To assess this influence, we numerically evolve the \emph{full} LME, in which the second term of Eq. \eqref{eq:lme} takes the form
\begin{align}
\gamma \sum_{i=1}^{N} \left( S_i^z \rho S_i^z - \frac{1}{2} \{(S_i^z)^2, \rho\} \right)
\end{align}
where $\gamma$ is the dephasing rate of the system. 
 This specific form of the LME models local dephasing for every site of both P1 and P2. By sweeping the dephasing rate \(\gamma\) and recording the peak end-to-end negativity, we obtain the curves in Fig. \ref{fig:fig8}.
For P2, we apply the TCG to obtain an effective dynamical equation for the density operator of the open system. The main difference compared to the TCG procedure used for the closed system is essentially the starting point itself, as we now begin from the LME instead of the VNE. Then, we obtain an effective master equation 
 \begin{align}
\dot{\rho}_{\text{eff}} = -i[H_{\text{eff}},\rho_{\text{eff}}]+\frac{\Gamma}{2} \sum_{j=e, r} (2S_j^z \rho_{\text{eff}} S_j^z - \{(S_j^z)^2, \rho_{\text{eff}}\}),
\end{align}
where 
\begin{align}
    \Gamma=\gamma\sum_k\frac{\bar{\lambda}^2_k}{\zeta^2}
    \label{eq:effdeprate}
\end{align}
and $H_{\text{eff}}$ is the same as for the closed system, Eq. \eqref{eq:Heff}. $\bar{\lambda}_k$ and $\zeta$ are defined in Eq. \eqref{eq:lambda} and Eq. \eqref{eq:zeta}, respectively.

Now, we can understand where P2's enhanced resilience to dephasing comes from.  The key difference lies in how the entanglement is generated: P1 requires the physical propagation of excitations through intermediate spins before boundary entanglement can be established. As each of these intermediate spins becomes populated, the system accumulates dephasing noise at each site. This sequential exposure results in a larger bulk population (see the red curve in the inset of Fig. \ref{fig:fig8}) and sharper entanglement degradation. In contrast, P2 consistently maintains low bulk occupation (orange curve in the inset of Fig. \ref{fig:fig8}) across all tested $\Delta/\delta$ ratios. This population suppression directly explains the significantly flatter negativity decay curve for P2 in the main panel of Fig. \ref{fig:fig8}: by avoiding the buildup of noise along the chain, it preserves the entanglement more effectively under dephasing.

The enhanced protection in P2 comes from two key factors related to the dimerization ratio $\Delta/\delta$. First, higher values push the system deeper into the dispersive regime, suppressing the real chain occupation. Second, the effective dephasing rate Eq. \eqref{eq:effdeprate} decreases quadratically with $\zeta_k$, explaining why P2's negativity curves in Fig. \ref{fig:fig8} decay increasingly slowly as $\Delta/\delta$ grows. The combination of these effects, i.e., the minimal bulk population and suppressed $\Gamma$, gives P2 its characteristically flat negativity decay.

In contrast, P1's excitation-mediated transport remains fundamentally exposed to dephasing regardless of $\Delta/\delta$, as its physical propagation mechanism inevitably populates intermediate sites. While stronger dimerization may slightly reduce bulk occupation, it cannot eliminate the accumulation of sequential noise along the chain. This stark difference highlights the central advantage of virtual tunneling: by avoiding real excitations in the bulk, P2 naturally decouples from noise sources while maintaining efficient end-to-end entanglement generation.

\section{NON-MARKOVIAN DYNAMICS} 

When dealing with non-Markovian dynamics, a more detailed characterization of the bath becomes unavoidable. This contrasts with the Markovian case, where the bath’s correlation functions are assumed to decay either instantaneously, as in the idealized Dirac $\delta$ limit, or on timescales much shorter than the system’s dynamics, thereby justifying the usage of a fast decaying spectral density. Such an approximation no longer holds in the non-Markovian regime, where the system retains memory of its past interactions with the environment. In this context, the central object is the bath’s spectral density function, $\gamma(\omega)$, which encodes both the distribution of bath frequencies and their coupling strength to the system.

To deduce the form of the spectral density, we must first define the bath's modes. A standard approach is to model the bath as a collection of non-interacting harmonic oscillators. The total Hamiltonian for the system and a bosonic reservoir is given by
\begin{align}
H_{\text{tot}} = H_S + \sum_k \omega_k b_k^\dagger b_k + \sum_k \left( g_k L b_k^\dagger + g_k^* L^\dagger b_k \right),
\end{align}
where $b_k$ and $b_k^\dagger$ are the annihilation and creation operators for the mode with frequency $\omega_k$, $L$ is a system operator, and $g_k$ are the coupling strengths. Assuming the bath remains in a thermal state, this model allows us to calculate the bath correlation functions analytically, whose properties are encapsulated in the spectral density $\gamma(\omega)$ \cite{Breuer1985}.

 Here we will use the pseudomode formalism \cite{Garraway1997}, which offers a conceptually clear and numerically efficient framework for simulating non-Markovian effects arising from reservoirs with Lorentzian or near-Lorentzian spectral densities.

The key insight of this method is that when the spectral density is Lorentzian,
\begin{align}
\gamma(\omega) = \frac{1}{\pi} \frac{g^{2} \kappa}{(\omega - \omega_{a})^{2} + \kappa^{2}},
\end{align}
the reservoir can be replaced by a single auxiliary damped harmonic oscillator (a pseudomode) with frequency $\omega_a$, coupled to the system with strength $g$, and itself damped at a rate $\kappa$. For our case, the total Hamiltonian of the combined system and pseudomode is
\begin{equation}
H_{\text{pm}} = H_S \otimes \mathbbm{1} + \mathbbm{1} \otimes \omega_a a^\dagger a + g \left( L \otimes a^\dagger + L^\dagger \otimes a \right),
\label{eq:Hpm}
\end{equation}
where $a$ and $a^\dagger$ are the annihilation and creation operators of the pseudomode and the system–reservoir coupling is implemented via the operator
$
L = \sum_{i=1}^{N} \left( S_i^{z} + S_i^{x} \right),
$
which accounts for both dephasing and dissipative processes in a zero-temperature bath. The evolution of the joint density operator $\rho(t)$ is governed by the Markovian master equation

\begin{equation}
\frac{d\rho}{dt} = -i[H_{\text{pm}}, \rho] + \kappa \left( a\rho a^\dagger - \frac{1}{2}\{a^\dagger a, \rho\} \right).
\label{eq:master}
\end{equation}
This representation effectively maps a non-Markovian open system problem onto an extended Markovian one, enabling the use of Lindblad-form master equations to simulate memory effects without explicitly dealing with integrodifferential equations or memory kernels. The parameter $\kappa$ determines the width of the Lorentzian spectral density and thus controls the bath correlation time: small $\kappa$ corresponds to strong non-Markovianity with long bath memory, while large $\kappa$ recovers the Markovian limit with rapid environmental decoherence. Recent work has extended this approach to multiple pseudomodes for complex spectral densities~\cite{PhysRevA.97.032110} and fermionic environments~\cite{PhysRevLett.123.090402}, demonstrating its versatility in modeling modern quantum devices where environmental memory effects are significant~\cite{PhysRevX.11.021064}.

We note that richer non-Markovian features may emerge in the long-time dynamics. However, these effects lie beyond the scope of the present analysis, which is focused on fast, on-demand entanglement generation.

\begin{figure*}[h]
  \centering
    \caption{
    Maximum normalized end-to-end negativity achieved within an
    evolution time equal to twice the optimal transfer time of the
    closed system, for both protocols under non-Markovian dissipation
    modeled via the pseudomode method.
    Each column corresponds to a different spectral width $\kappa$ of
    the Lorentzian reservoir:
    (left) $\kappa = 0.01$ (strongly non-Markovian), (center) $\kappa =
    1$ (intermediate memory), and (right) $\kappa = 100$ (Markovian
    limit).
    The top row displays results for P1, and the bottom row for P2.
    For small $g$ and $\kappa$, P2 sustains high entanglement over a
    broader parameter region compared to P1.
    As $g$ increases, both protocols deteriorate, but P2 remains more
    resilient.
    In the Markovian limit, both recover the Lindblad results,
    validating the approach.
  }
  \includegraphics[width=\textwidth]{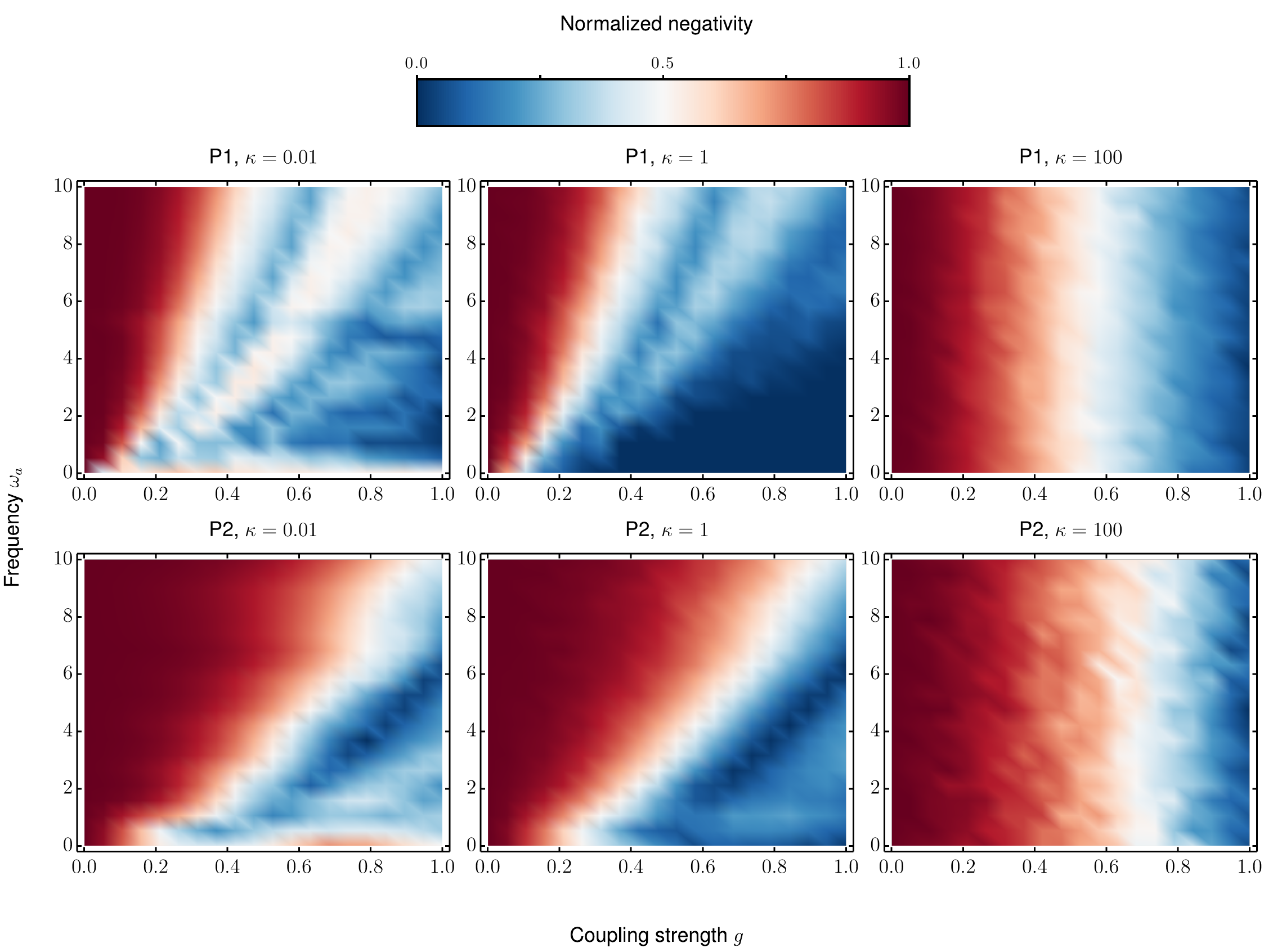}
  \caption*{\small{Source: The author.}}
  \label{fig:heatmaps}
\end{figure*}

Fig.\ref{fig:heatmaps} presents heatmaps of the maximum normalized end-to-end negativity achieved within an evolution time equal to twice the optimal entanglement generation time of the corresponding closed system, as a function of the system--reservoir coupling strength $\gamma$ and the environmental correlation time $\tau_c$.
\section{VALIDITY OF THE LINDBLAD MASTER EQUATION}

Throughout the derivation of the LME, an effort was made to highlight the restrictive scope of this master equation. In this section, we summarize some observations made during this work about the LME that will be of interest to anyone wishing to understand whether this master equation yields exact, approximate, or invalid results for a particular system. In this specific section and only here, we explicitly include $\hbar$ in equations, since it is necessary for comparisons with physically meaningful quantities.

Higher-order corrections in Eq. \eqref{eq:perturbativeser} involve an increasing number of time integrals. These integrals can be estimated \cite{stefanini2025lindbladme}, leading to a scaling of the form
\begin{align}
\sim \frac{\alpha}{\hbar} \left( \frac{\alpha \tau_{corr}}{\hbar} \right)^{n-1},
\end{align}
where $n$ denotes the order of the correction. The perturbative series converges provided that $\alpha \tau_{corr} / \hbar \ll 1$, which is consistent with the weak-coupling assumption. This allows us to determine that the combination of the coupling between system and bath and the timescale of the bath correlations must be such that $\alpha \tau_{corr} \ll \hbar$. 

A key element in the description of open quantum systems is the Fourier transform of the bath correlation functions, $\gamma(\omega)$, commonly referred to as the spectral density. A fundamental assumption in this dissertation is that the environment can be modeled as a large reservoir, with energy levels forming a continuous spectrum. This is crucial: if the environment were small, information about the system’s past states could feed back into its dynamics, generating memory effects and thereby breaking Markovianity.  

In most treatments using the Lindblad master equation, it is often assumed that $\gamma(\omega) = \text{const.}$, corresponding to a bath correlation function proportional to a Dirac delta, i.e., a function with vanishing correlation time $\tau_\text{corr} \to 0$. In this idealized limit, the Markovian approximation becomes formally exact. A situation that approximates this flat-spectrum limit arises when a quantum system is driven by a classical noisy field, for instance, the intensity or phase fluctuations of a laser, where the noise can be treated as effectively memoryless \cite{stefanini2025lindbladme}. In such cases, $\gamma(t) \propto \delta(t)$ provides a convenient idealization, yielding perfectly Markovian dynamics. However, in any realistic physical system, the bath always has a finite correlation time, no matter how short, which inevitably introduces non-Markovian corrections.

The rotating wave approximation relies on a separation of timescales: fast oscillations average out. It works when transitions are fully non-degenerate or degenerate. For nearly-degenerate transitions, these oscillations are slow, so you cannot neglect cross terms, and the standard Lindblad equation derived here fails to capture the dynamics correctly. This highlights, once again, that the rotating wave approximation removes valid physical process, thus being possible for the LME to take us to a completely different scenario from the one happening in real life if the proper care is not taken.

Sometimes, the assumption that $\alpha \ll 1$ can't be made, for an example see the Kondo model \cite{hewson1997kondo}, where for low temperatures the bath correlation time is divergent, indicating that this system needs the strongly interaction assumption to be well described, making it impossible to study it under any finite-order perturbative approach. 

In this dissertation, we applied the LME to study dephasing in a spin chain, a paradigmatic many-body quantum system. In general, the dissipators for a many-body system do not simply reduce to a sum of local dissipators acting independently on each subsystem, because the eigenstates of an interacting system have support across multiple subsystems. Consequently, the jump operators inherit this nonlocal character. Although this may seem counterintuitive, since system-bath interactions are often local, it is dictated by the requirement of thermalization: the jump operators must "know" the correct eigenstates of the full system Hamiltonian to drive it toward the correct Gibbs state as $t \rightarrow \infty$. Indeed, a master equation that effectively relaxes to the correct Gibbs state can be constructed for spin chains, indicating that experimentally observed behavior can be approached using the LME \cite{guimaraes2016nonequilibrium,silva2023nontrivial, landi2022nonequilibrium}. In our case, however, we do not need to address this full complexity due to the short time scales of the dynamics under consideration. Within this short-time regime, dephasing can be modeled phenomenologically by treating the $S^z$ operator as acting locally on the positions of the chain, effectively capturing the experimentally observed behavior. This approximation is valid at early times, as employed here, while equilibration with the environment, which would involve the full nonlocal structure of the jump operators, occurs only at later times and is not accounted for in this specific phenomenological approach.

The main takeaway from this section is that one should not apply the LME naively beyond its range of validity. While doing so may be tempting due to the simplicity of expression Eq. \eqref{eq:lme}, it is important to remember that obtaining a mathematically valid quantum state at the end of a simulation does not guarantee that the result is the physically accurate one.
\chapter{CONCLUSIONS}

\vspace{0.8cm}
The purpose of this work was to investigate, in a systematic and in-depth manner, the generation of entanglement in XX spin chains, considering both ideal scenarios and realistic situations in which the effects of disorder and coupling to the environment cannot be neglected. This objective is part of the broader context of quantum information processing and communication, where entanglement plays a central role as a fundamental resource for teleportation protocols, quantum key distribution, and distributed computation. Throughout the dissertation, two distinct entanglement generation protocols, referred to as P1 and P2, were analyzed, contrasting their architectures, dynamical mechanisms, and resilience to different sources of noise.

The study revealed that, although both protocols are capable of generating end-to-end entanglement in finite chains, the performance of protocol P2 consistently stands out in all aspects investigated. In particular, it was found that P2 reaches higher values of negativity in shorter evolution times than P1, both in spin-1/2 chains and in chains with higher spin magnitudes (s = 1 and s = 3/2). This efficiency gain can be attributed to its architecture, which exploits optimized boundary magnetic fields to virtually couple the end spins while minimizing real excitations of the intermediate spins. This feature is crucial, as it reduces the vulnerability of the protocol to imperfections inherent in real solid-state systems.

Another fundamental point concerns robustness against disorder. In chains subject to diagonal perturbations (random local fields) or off-diagonal perturbations (fluctuations in coupling constants), it was observed that P2 sustains higher average entanglement values and exhibits smaller dispersions compared to P1. This means that, in addition to being more efficient, protocol P2 is also more predictable and stable under adverse experimental conditions, making it a more promising candidate for implementation in solid-state quantum devices.

The analysis was subsequently extended to the regime of open quantum systems, in which the chains interact with a dissipative environment. In particular, the case of dephasing, modeled by the LME, was considered. Numerical results show that the entanglement generated by P2 decays significantly more slowly than that produced by P1. This difference can be explained by the fact that, while P1 depends on the real propagation of excitations through the chain—accumulating noise at each intermediate site; protocol P2 keeps the channel spins practically inactive, so that the degradation of the entangled state occurs at a much milder rate. This result reinforces the idea that strategies based on effective couplings and virtual excitations provide natural protection against dissipative processes.

From a conceptual and methodological perspective, the dissertation also presented relevant contributions. The use of effective models, such as the trimer approximation in the strong dimerization regime and the effective Hamiltonian in the dispersive regime of P2, provided an analytical basis for understanding the mechanisms underlying robust entanglement generation. These simplified models not only help interpret the numerical results but also point to avenues for engineering new spin chain architectures capable of optimizing resilience against noise.

It is important, however, to acknowledge the limitations of the present study. The simulations were primarily carried out on relatively small chains to reduce computational complexity. While these choices are justifiable and relevant for several experimental platforms, the use of the LME imposes additional constraints: it relies on the assumptions of weak system-environment coupling, Markovianity, and the rotating-wave approximation, which do not fully capture general non-Markovian effects, strong interactions, or structured environmental spectra. A complete and more faithful description of realistic systems would therefore require future investigations that explore longer chains, long-range interactions, time-dependent noise, and more complex non-Markovian environments.

Nevertheless, the results presented here allow for clear directions to be outlined for future work. Among them, we highlight: (i) extending the analyses to larger chains to assess the scalability of the protocols; (ii) studying environments with nontrivial spectra to understand the limits of P2’s protection; (iii) investigating hybrid architectures in which spin chains are integrated with bosonic or fermionic modes, (iv) Implement these protocols using alternative master equations that capture effects beyond the scope of the LME and (v) We emphasize that our exploration of non-Markovian effects is a preliminary step, carried out under the assumption of a simple yet physically relevant Lorentzian spectral density. More intricate environments—such as sub-ohmic, super-ohmic, or multi-peaked spectra \cite{Wilner2015SuperOhmicSubOhmic}—would require additional pseudomodes to faithfully reproduce their structure and could give rise to qualitatively new dynamical behaviors. 

Finally, it is worth emphasizing the experimental relevance of the results obtained. The properties of protocol P2, particularly its scalability, speed, and resilience to imperfections, make it a viable scheme for implementation in several well-established platforms, such as superconducting qubits in circuit QED architectures, trapped ions with phononic interactions, and NV centers in diamond. These technologies already offer local control of fields and couplings with the precision required to reproduce the conditions analyzed theoretically in this dissertation.

This work has shown that SCs constitute a promising platform for the rapid and robust generation of entanglement, provided they are properly designed. The systematic comparison between protocols P1 and P2 clearly demonstrated that appropriate architectural choices are decisive for ensuring efficiency and robustness against noise, and that such choices can be grounded in both numerical analyses and effective models. Thus, this dissertation contributes not only to advancing the fundamental understanding of entanglement dynamics in spin chains but also to opening concrete pathways for their application in emerging quantum technologies, consolidating itself as a relevant step in the effort to bring theory, practice and technology closer together in the field of quantum information.

The work presented in this dissertation served as base for a paper:
\begin{itemize}
\item Comparative Analysis of Robust Entanglement Generation in
Engineered XX Spin Chains, available on the journal Entropy, \href{https://www.mdpi.com/1099-4300/27/7/764}{https://www.mdpi.com/1099-4300/27/7/764}.
\end{itemize}

%
\phantompart
\postextual
\backmatter
\renewcommand{\bibname}{\normalsize REFERENCES \setlength{\afterchapskip}{\baselineskip}}
\bibliography{bibliography.bib}

\begin{apendicesenv}
\apendices
\chapter{COMPLEMENTS} \label{apendB}
\vspace{0.8cm}
\textbf{COMPLETE METRIC SPACES}

A complete metric space is a space where every Cauchy sequence converges. A Cauchy sequence is a sequence of elements \( \{a_n\} \) such that for every positive real number \( \epsilon \), there exists a positive integer \( N \) where, for all natural numbers \( m, n > N \),
\[
\|a_m - a_n\| < \epsilon.
\]
In other words, the terms of the sequence get arbitrarily close to each other after a finite number of steps. An example of a Cauchy sequence is \( a_n = \frac{1}{n} \).
\begin{figure}[h]
    \centering
    \caption{Cauchy sequence for  \( a_n = \frac{1}{n} \) for the interval [0,100].}
    \includegraphics[width=0.8\linewidth]{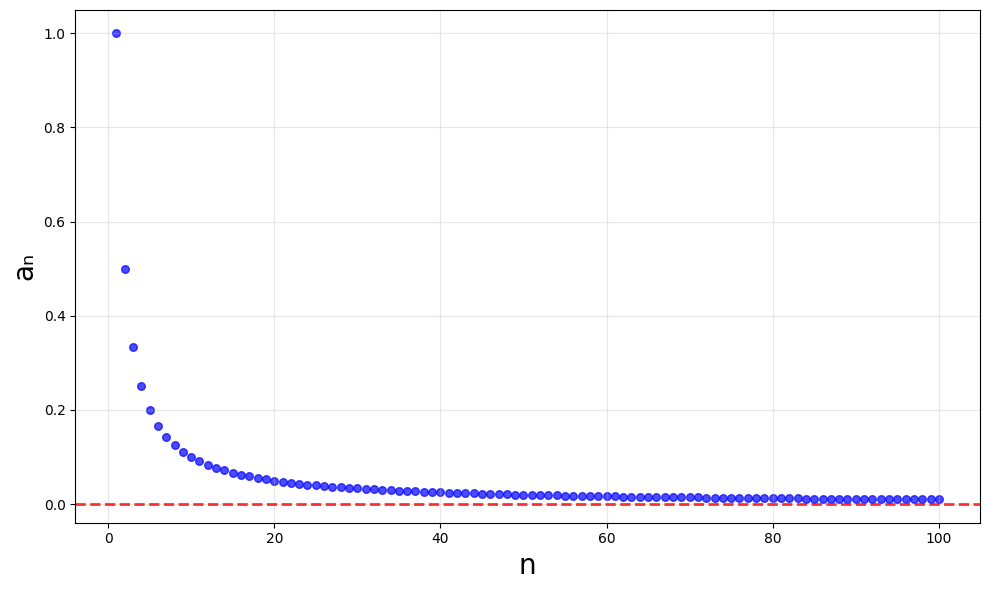}
    \caption*{\small{Source: The author.}}
    \label{fig:cauchyseq}
\end{figure}

\section*{\textbf{BLOCH SPHERE}}
The Bloch sphere provides a powerful geometrical representation for the pure state space of a two-level quantum system, commonly known as a qubit. This elegant representation, named after the Swiss-American physicist Felix Bloch, offers an intuitive visualization of quantum states and their transformations.

The foundation of this representation stems from the fundamental quantum mechanical postulate requiring state vectors to be normalized. For any pure state $\ket{\psi}$ of a qubit, this normalization condition reads:
\begin{align}
\langle \psi | \psi \rangle = 1
\end{align}
This constraint permits a general parameterization of qubit states using spherical coordinates. The most general pure state of a two-level system can be expressed as:
\begin{align}
\ket{\psi}=\cos\left(\frac{\theta}{2}\right)\ket{0}+e^{i\phi}\sin\left(\frac{\theta}{2}\right)\ket{1}
\end{align}
where ${\ket{0}, \ket{1}}$ forms an orthonormal basis for the two-dimensional Hilbert space. To represent a state in the Bloch sphere we set $\theta$ as the polar angle, and $\phi$ as the azimuthal angle.
The geometric interpretation of these parameters leads naturally to the Bloch sphere representation. Each pure qubit state corresponds to a unique point on the surface of a unit sphere in $\mathbb{R}^3$, with:

\begin{itemize}
\item The north pole ($\theta = 0$) representing the $\ket{0}$ state
\item The south pole ($\theta = \pi$) representing the $\ket{1}$ state
\item Points along the equator ($\theta = \pi/2$) representing all possible equal superpositions of $\ket{0}$ and $\ket{1}$
\end{itemize}

The spherical symmetry of this representation reflects the fact that all pure states are equivalent up to unitary transformations.

\begin{figure}[H]
    \centering
    \caption{A state represented in the Bloch sphere.}
    \includegraphics[width=0.5\linewidth]{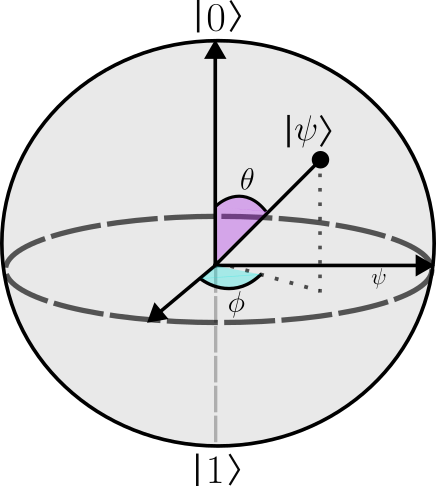}
    \caption*{\small{Source: The author.}}
    \label{fig:placeholder}
\end{figure}

\section*{\textbf{JORDAN-WIGNER TRANSFORMATION}}

The study of one-dimensional quantum spin chains begins with the Heisenberg model, which provides a fundamental framework for understanding magnetic systems. To analyze this system further, it becomes convenient to express the Hamiltonian in terms of spin raising and lowering operators through the relations,
\begin{align}
    \sigma_x^i &= \sigma_+^i + \sigma_-^i, \\
    \sigma_y^i &= \frac{\sigma_+^i - \sigma_-^i}{i}.
\end{align}
Additionally, by parameterizing the coupling constants as $J_x=J\left(\frac{1+\gamma}{2}\right)$ and $J_y=J\left(\frac{1-\gamma}{2}\right)$, the Hamiltonian transforms into
\begin{align}
    H = -\sum_{i=1}^L \Bigl\{ 
    &J(\sigma^+_i\sigma^-_{i+1} + \sigma^-_i\sigma^+_{i+1}) + J\gamma(\sigma^+_i\sigma^+_{i+1} + \sigma^-_i\sigma^-_{i+1}) + 2g\sigma^+_i\sigma^-_i 
    \Bigr\}.
\end{align}

This form reveals a striking resemblance to the tight-binding Hamiltonian in second quantization
\begin{align}
    H = -t \sum_{\langle i,j\rangle,\sigma} (c_{i,\sigma}^{\dagger} c_{j,\sigma} + \text{h.c.}),
\end{align}
which describes fermionic hopping between lattice sites \cite{Ashcroft1976} However, despite this similarity, a crucial distinction lies in their operator algebras. While fermionic operators satisfy $\{c_i,c^{\dagger}_j\}=\delta_{ij}$, Pauli operators exhibit mixed commutation relations: $[\sigma^i_-,\sigma^j_+]=0$ for $i\neq j$ and $\{\sigma^i_-,\sigma^i_+\}=1$ for $i=j$, demonstrating their neither purely fermionic nor bosonic nature.

The Jordan-Wigner transformation \cite{landiJordan} provides a mapping between spin operators and fermionic creation/annihilation operators. This non-local transformation is defined as
\begin{align}
    c_i = \bigg[\prod_{n=1}^{i-1}\bigl(-\sigma_z^n\bigr)\bigg]\sigma_-^i,
\end{align}
which can be visualized as
\begin{align}
    c_i=(-\sigma_z)\otimes(-\sigma_z)\otimes\dots\otimes(-\sigma_z) \otimes\sigma^i_-\otimes(\mathbb{I})\otimes(\mathbb{I})\otimes\dots\otimes(\mathbb{I}).
\end{align}
The sequence of $(-\sigma_z)$ operators, known as the Jordan-Wigner string, ensures the proper fermionic behavior.

The validity of this transformation becomes apparent when examining the resulting operator algebra,
\begin{align}
    c_{i} c_{i}^{\dagger} &= \left[ \prod_{n=1}^{i-1} (-\sigma_{z}^{n}) \right] \sigma_{-}^{i} \left[ \prod_{m=1}^{i-1} (-\sigma_{z}^{m}) \right] \sigma_{+}^{i} = \sigma^i_-\sigma^i_+,
\end{align}
and similarly for $c^{\dagger}_ic_i=\sigma^i_+\sigma^i_-$, yields the expected fermionic relation $\{c_i,c^{\dagger}_i\}=1$. For the case of $i\neq j$ (assuming $j>i$), the transformation yields
\begin{align}
    c_i c_j^\dagger &= \bigg[\prod_{n=1}^{i-1}(-\sigma_z^n)\bigg]\sigma_-^i\bigg[\prod_{m=1}^{j-1}(-\sigma_z^m)\bigg]\sigma_+^j \\
    &= -\left[ \prod_{n=i}^{j-1} \left(-\sigma_{z}^{n}\right) \right] \sigma_{-}^{i} \sigma_{+}^{j},
\end{align}
where the non-local nature of the transformation introduces the sign change through the commutation relations of Pauli matrices
\begin{align*}
    \sigma_+(-\sigma_z) &= \sigma_+, & \sigma_-(-\sigma_z) &= -\sigma_-, \\
    (-\sigma_z)\sigma_+ &= -\sigma_+, & (-\sigma_z)\sigma_- &= \sigma_-.
\end{align*}

When combined with $c_j^\dagger c_i$, this results in the vanishing anti-commutator for $i\neq j$, thereby preserving the fermionic algebra. This mathematical framework demonstrates that exchange interaction problems described by the Heisenberg Hamiltonian can be equivalently formulated as fermionic hopping problems, and vice versa, through the Jordan-Wigner transformation.
\end{apendicesenv}

\end{document}